\documentclass[aps,prc,preprint,superscriptaddress,showpacs,amsfonts,floatfix,nofootinbib]{revtex4}
\usepackage{graphicx}
\usepackage{epsfig}
\usepackage{amsmath}
\usepackage{wasysym}

\usepackage[usenames]{color}
\usepackage{epsfig}
\usepackage{epstopdf}
\usepackage{amssymb,amsmath}
\usepackage{amsmath}
\usepackage{epstopdf}
\usepackage{sidecap}
\usepackage{cases}
\usepackage{enumitem}
\usepackage{bm}
\usepackage{overpic}
\usepackage{upgreek}
\usepackage{morefloats}
\usepackage{hyperref}

\begin{document}

\title{The complete experiment for
photoproduction of pseudoscalar mesons in a truncated partial wave analysis}
\author{Y. Wunderlich}
\author{R. Beck}
\affiliation{Helmholtz-Institut f\"ur Strahlen- und Kernphysik, Universit\"at Bonn, Bonn, Germany}
\author{L.~Tiator}
\affiliation{Institut f\"ur Kernphysik, Johannes
Gutenberg-Universit\"at Mainz, Mainz, Germany}

\date{February 21, 2014}
\begin{abstract}
The complete experiment problem in the truncated partial wave analysis of pseudoscalar meson
photoproduction with suppressed t-channel exchanges is investigated. The focus is set to
ambiguities of the group S observables with the unpolarized differential cross section, $\sigma_0$,
and the three single-spin observables, $\Sigma$, $T$ and $P$. For this purpose, the approach and
formalism already worked out by Omelaenko in 1981 is revisited in this work. A numerical study
using multipoles of the PWA solution MAID2007 shows how only one additional double polarization
observable can resolve all ambiguities. Therefore, the possibility emerges to  perform a complete
experiment with only five observables.
\end{abstract}

\maketitle

\section{Introduction} \label{sec:Introduction}

The nucleon and its excitation spectrum is of fundamental interest
for our understanding of the visible nature in terms of quantum
chromodynamics (QCD) in the non-perturbative regime. Whereas the
nucleon itself is mainly investigated in electron scattering by its
form factors and densities as well as in Compton scattering by
polarizabilities, the excitation spectrum is traditionally explored
in elastic and inelastic pion nucleon scattering and meson photo-
and electroproduction. While the electromagnetic excitation of
nucleon resonances was for a long time just the source for obtaining
the photon decay amplitudes and the transition form factors, in
recent years, the accuracy of data in photo- and electroproduction
has increased so much that this reaction has now also become a
source for possible observations of new resonances or for
confirmations and establishments of such resonances that have only
been `seen' in other reactions with rather uncertain parameters in
the Particle Data Listings. Just recently in the 2012 issue of the
listings of the Particle Data Group (PDG) a series of $N^*$
resonances have been established mainly due to precise data in kaon
photoproduction~\cite{Beringer:1900zz,Anisovich:2011fc}.

The simplest process to detect and to study nucleon resonances is
the elastic pion nucleon scattering. It has the largest cross
sections, it is a two-body process with a simple kinematical
structure and it is described by only two spin degrees of freedom,
giving rise to two scattering amplitudes and four polarization
observables. This field was pioneered by Hoehler\cite{Hoehler84} and
Cutkosky\cite{CMB} and led to the detection of most of the $N^*$ and
$\Delta$ resonances. Their determinations of masses, widths, partial
decay widths, pole positions and residues are still considered as of
high quality in the PDG. After shutdown of the pion beams,
experimental activities in pion nucleon scattering practically
stopped about 20 years ago. Nevertheless, an impressive progress has
been achieved in the last decade, mostly by shaping up the analyzing
tools and developments of various models, first to mention the
dynamical models, some of them with 8 and more coupled
channels~\cite{Arndt:2006bf,Drechsel:2007if,Chen:2007cy,Ronchen:2012eg,
Kamano:2013iva,Shrestha:2012ep,Shklyar:2012js}.

On the other side, the construction of modern electron accelerators,
new detector systems and polarized targets led to an enormous
progress on experiments in photo- and electroproduction. Next to
pion nucleon scattering, the photoproduction of pseudoscalar mesons,
$(\pi, \eta, \eta', K)$ is the simplest process to analyze. It is
described by four spin degrees of freedom with 4 complex amplitudes,
usually given as CGLN, invariant, helicity or transversity
amplitudes, all of them are linearly related to each other. With
these four amplitudes, 16 polarization observables are defined and
can be measured with linearly or circularly polarized photon beams,
polarized targets and recoil polarization detection.

Already around the year 1970 people started to think about how to determine the four complex
helicity amplitudes for pseudoscalar meson photoproduction from a complete set of experiments. In
1975 Barker, Donnachie and Storrow published their classical paper on `Complete
Experiments'~\cite{Barker75}. After reconsiderations and careful studies of discrete ambiguities,
in the 90s~\cite{Keaton:1996pe,chiang} it became clear that such a model independent amplitude
analysis would require at least 8 polarization observables (including the unpolarized cross
section) which have to be carefully chosen. There are a large number of possible combinations, but
all of them would require a polarized beam and target and in addition also recoil polarization
measurements. Technically this was not possible until very recently, when transversely polarized
targets came into operation at Mainz, Bonn and JLab and furthermore recoil polarization
measurements by nucleon rescattering have been shown to be doable.

A complete experiment is a set of measurements which is sufficient
to predict all other possible experiments, provided that the
measurements are free of uncertainties. Therefore it is first of all
an academic problem, which can be solved by mathematical algorithms.
In practise, however, it will not work in the same way and either a
very high statistical precision would be required, which is very
unlikely, or further measurements of other polarization observables
are necessary. This has been studied by
Ireland~\cite{Ireland:2010bi} with information entropy, by a joint
Mainz/GWU collaboration~\cite{Workman:2011hi} with event based
pseudo data generated from the MAID model~\cite{Drechsel:2007if}, by
a JLab collaboration with both experimental and pseudo data for kaon
photoproduction~\cite{Sandorfi:2010uv} and in a very recent work by
the Ghent group~\cite{Vrancx:2013pza} with a combination of kaon
photoproduction data measured at GRAAL and additional pseudo data
from a theoretical model. In fact, photoproduction of $K\Lambda$ and
$K\Sigma$ are ideal for the complete experiment analysis, as the
necessary recoil polarization observables can be obtained from the
self-analyzing decay of the hyperons. In case of pion and eta
photoproduction this is very different and recoil polarization can
only be detected by an additional elastic scattering of the outgoing
nucleon on a spin-zero nucleus as $^{12}C$~\cite{Sikora:2013vfa}.
This reduces already very much the count rates, but even more, it
does only allow a measurement of the transverse component of the
recoil polarization in the laboratory frame. In this way, the
necessary recoil polarization observables in the CMS frame cannot be
measured.

But even for kaon photoproduction, where the first complete experiment analysis is only a question
of time, an important problem remains with the unknown overall phase. Any set of quadratic
equations must suffer from the problem that the underlying amplitudes can only be solved up to an
overall phase. For the four complex amplitudes in pseudoscalar photoproduction, this means, that
the full solution gives just 4 absolute magnitudes and 3 relative phases. The residual overall
phase remains undetermined. In the literature, two methods have been discussed, which are both
highly academic and cannot be used in practise. The first goes back to
Goldberger~\cite{Goldberger:1963} in 1963 with a Hanbury-Brown and Twiss experiment, the second was
recently published by Ivanov~\cite{Ivanov:2012na} in 2012, using vortex beams to measure the phase
of a scattering amplitude. Even though the missing overall phase is no problem for reconstructing
all 16 possible polarization observables, it does not allow to perform a partial wave expansion,
because of the fact that this phase is a function of both energy and
angle~\cite{Tiator:2011tu,Tiator:2011a}. Nevertheless, if the complete experiment can be performed,
it will be the optimal condition for a partial wave analysis.


In order to obtain the partial wave amplitudes and subsequently the information on nucleon
resonances, another approach has to be undertaken, the Truncated Partial Wave Analysis (TPWA). In
this method, all 16 polarization observables are expanded in a partial wave series up to a given
maximal angular momentum $\ell_{max}$, where all partial wave amplitudes are only functions of the
energy. In 1981 Omelaenko~\cite{omel} showed that such a complete truncated partial wave analysis
is possible with even less than 8 observables. In fact he proved that with only 4 observables,
unpolarized cross section $\sigma_0$, photon beam asymmetry $\Sigma$, target polarization $T$ and
recoil polarization $P$, the sets of quadratic equations with multipoles can be solved up to a
discrete ambiguity for any given $\ell_{max}$. And in order to resolve this final ambiguity, only
one more double polarization is needed, e.g. $F,G,C_{x'},O_{x'},C_{z'},O_{z'}$, while a measurement
of $E$ or $H$ would not suffice. This is a rather surprising result, as it even allows a complete
analysis for pion or eta photoproduction without the need of recoil polarization observables. The
single recoil polarization $P$ can more easily be measured in a beam-target double polarization
experiment.

As in the previous case, also here, the full solution will determine
all partial waves only up to an overall phase, however, this phase
is now only dependent on the energy, and with some theoretical
assumptions, e.g. unitarity and Watson theorem, this phase can be
constructed. This was first performed for $\ell_{max}=1$ in 1989 by
Grushin et al.~\cite{grushin} for a complete TPWA in the Delta
region.

The aim of this paper is to revisit the Omelaenko paper~\cite{omel},
published more than 30 years ago. The formalism of this paper is not
so easy to follow in the shortness of the original publication and
the paper never gained much attention. We have extended and further
clarified the formalism and have applied the method of ambiguities
to modern partial wave analyses (PWA) as MAID~\cite{MAID},
SAID~\cite{SAID} and BnGa~\cite{BoGa}. Furthermore, we have also
considered truncations beyond $S+P$ waves and discuss also higher
partial waves. We also investigate the possibilities for unique
numerical solutions with current PWA.

The work of Omelaenko is based on investigations on ambiguities
arising in the analysis of $\pi N$ scattering that were performed by
Gersten~\cite{Gersten} in 1969. Both approaches proceed via
appropriately representing the spin amplitudes describing the
process by products. For the sake of completeness, it should also be
mentioned that for $\pi N$ scattering an alternative scheme for
obtaining product representations was proposed by
Barrelet~\cite{Barrelet} in 1972 (see~\cite{VanHorn} for a brief
treatment on this subject). The latter approach is generally
referred to as the method of Barrelet zeros.

After a general introduction to the basics of the pseudoscalar meson
photoproduction process, in Sec. 3 we derive the ambiguities of the
group S observables (unpolarized cross section $\sigma_{0}$,
photon beam asymmetry $\Sigma$, target asymmetry $T$ and nucleon
recoil polarization $P$) for reconstructing e.m. multipoles
following the method of Omelaenko. In Sec. 4 we discuss the
behavior of double-polarization observables and their ability to
resolve ambiguities in the partial wave solutions. In Sec. 5 we
present a detailed study of the example with $\ell_{max}=1$. At the
end we give a short summary and an outlook for applications with
experimental data in the near future. In an appendix we finally
collect somewhat lengthy but useful mathematical formalism.

\section{Basic definitions} \label{sec:Definitions}

For photoproduction of pseudoscalar mesons on the nucleon,
\begin{equation}
\gamma N \rightarrow \varphi B \,, \label{eq:PhotProdProcess}
\end{equation}
where $\varphi$ denotes the pseudoscalar meson and $B$ the recoil
baryon in the final state, the amplitude can be written in a general
form~\cite{CGLN}
\begin{equation}
\mathcal{F} = \chi_{m_{s_{f}}}^{\dagger} F_{\mathrm{CGLN}} \hspace*{2pt} \chi_{m_{s_{i}}} \mathrm{.} \label{eq:FullAmplitude}
\end{equation}
The spinors $\chi_{m_{s_{i}}}$ and $\chi_{m_{s_{f}}}$ describe the
initial nucleon as well as the recoil baryon in the final state. The
spin operator $F_{\mathrm{CGLN}}$ appearing in
Eq.~(\ref{eq:FullAmplitude}) has the following expansion into spin
momentum terms~\cite{CGLN}
\begin{equation}
  F_{\mathrm{CGLN}} = i  \vec{\sigma} \cdot \hat{\epsilon}\; F_{1} + \vec{\sigma} \cdot \hat{q}
  \; \vec{\sigma} \cdot \hat{k} \times \hat{\epsilon}\; F_{2} + i  \vec{\sigma} \cdot \hat{k}\; \hat{q}
  \cdot \hat{\epsilon}\; F_{3} + i \vec{\sigma} \cdot \hat{q} \; \hat{q} \cdot \hat{\epsilon}\; F_{4}\,.
  \label{eq:DefCGLN}
 \end{equation}

In Eq.~(\ref{eq:DefCGLN}), $\hat{\epsilon}$ denotes the polarization
unit vector of the incoming photon and $\hat{k} = \vec{k} / \left|
\vec{k} \right|$ as well as $\hat{q} = \vec{q} / \left| \vec{q}
\right|$ are the normalized 3-momenta of the incoming and outgoing
particles in the center of mass system (CMS). The complex coefficients $\left\{ F_{i}
\left(W, \hspace*{2pt} \theta\right), i = 1,\ldots,4 \right\}$,
carrying dependencies on the total CMS energy $W$ and the CMS
scattering angle $\theta$ are called CGLN amplitudes (abbreviation for Chew, Goldberger, Low and Nambu). Once they are
known, the photoproduction process is described completely. The
angular dependence of the $F_{i} \left(W, \hspace*{2pt}
\theta\right) $ is given in terms of the multipole
expansion~\cite{CGLN,Sandorfi:2010uv}.
\begin{align}
F_{1} \left( W, \theta \right) &= \sum \limits_{\ell = 0}^{\infty} \Big\{ \left[ \ell M_{\ell+} \left( W \right) + E_{\ell+} \left( W \right) \right] P_{\ell+1}^{'} \left( \cos \theta \right) \nonumber \\
 & \quad \quad \quad + \left[ \left( \ell+1 \right) M_{\ell-} \left( W \right) + E_{\ell-} \left( W \right) \right] 
  P_{\ell-1}^{'} \left( \cos \theta \right) \Big\} \mathrm{,} \label{eq:MultExpF1} \\
F_{2} \left( W, \theta \right) &= \sum \limits_{\ell = 1}^{\infty} \left[ \left( \ell+1 \right) M_{\ell+} \left( W \right) + \ell M_{\ell-} \left( W \right) \right] 
  P_{\ell}^{'} \left( \cos \theta \right) \mathrm{,} \label{eq:MultExpF2} \\
F_{3} \left( W, \theta \right) &= \sum \limits_{\ell = 1}^{\infty} \Big\{ \left[ E_{\ell+} \left( W \right) - M_{\ell+} \left( W \right) \right] P_{\ell+1}^{''} \left( \cos \theta \right) \nonumber \\
 & \quad \quad \quad + \left[ E_{\ell-} \left( W \right) + M_{\ell-} \left( W \right) \right] P_{\ell-1}^{''} \left( \cos \theta \right) \Big\} \mathrm{,} \label{eq:MultExpF3} \\
F_{4} \left( W, \theta \right) &= \sum \limits_{\ell = 2}^{\infty} [ M_{\ell+} \left( W \right) - E_{\ell+} \left( W \right) - M_{\ell-} \left( W \right) 
  - E_{\ell-} \left( W \right) ] P_{\ell}^{''} \left( \cos \theta \right) \mathrm{,} \label{eq:MultExpF4}
\end{align}
where the electric and magnetic multipoles $E_{\ell\pm}$ and
$M_{\ell\pm}$ describe transitions induced by electric and magnetic
photons, respectively. The summation index $\ell$ quantizes the
orbital angular momentum of the final $\varphi B$ system, which has
a total angular momentum $J = \ell \pm 1/2$, and $P_{\ell} \left(
\cos \theta \right)$ are the Legendre polynomials.

For certain photoproduction channels ($\gamma p \rightarrow \pi^{0}
p$ is an important example but $\gamma p \rightarrow \eta p$ is also
applicable), close to production thresholds and in the low energy
region, a truncation of the infinite series (\ref{eq:MultExpF1}) to
(\ref{eq:MultExpF4}) at a finite value $\ell_{\mathrm{max}} = L$
already yields a good approximation for the
$F_{i}$~\cite{Sandorfi:2010uv}. Those channels are at the center of
attention in this work. Besides the CGLN amplitudes $F_{i}$, also
other sets of amplitudes, helicity, transversity and invariant
amplitudes are commonly used. The transversity amplitudes $\left\{
b_{i} \left(W, \hspace*{2pt} \theta\right), i = 1,\ldots,4 \right\}$
are defined by a rotation of the spin quantization axis of the
target nucleon and recoil baryon to the normal of the reaction
plane~\cite{Barker75,FTS92}
\begin{align}
 b_{1} \left( W, \theta\right) &= - b_{3} \left( W, \theta\right)
  + i \mathcal{C} \sin \theta \left[ F_{3} \left( W, \theta\right) e^{- i \frac{\theta}{2}} + F_{4} \left( W, \theta\right) e^{ i \frac{\theta}{2}} \right] \mathrm{,} \label{eq:b1BasicForm} \\
 b_{2} \left( W, \theta\right) &= - b_{4} \left( W, \theta\right)
  - i \mathcal{C} \sin \theta \left[ F_{3} \left( W, \theta\right) e^{i \frac{\theta}{2}} + F_{4} \left( W, \theta\right) e^{- i \frac{\theta}{2}} \right] \mathrm{,} \label{eq:b2BasicForm} \\
 b_{3} \left( W, \theta\right) &= \mathcal{C} \left[ F_{1} \left( W, \theta\right) e^{- i \frac{\theta}{2}} -  F_{2} \left( W, \theta\right) e^{ i \frac{\theta}{2}} \right] \mathrm{,} \label{eq:b3BasicForm} \\
 b_{4} \left( W, \theta\right) &= \mathcal{C} \left[ F_{1} \left( W, \theta\right) e^{i \frac{\theta}{2}} -  F_{2} \left( W, \theta\right) e^{- i \frac{\theta}{2}} \right] \mathrm{.} \label{eq:b4BasicForm}
\end{align}
In the following, we will drop the $W$ dependence of the amplitudes
and all further considerations and analyses will be single-energy analyses,
where the energy $W$ is kept fixed. $\mathcal{C}$ is a
complex factor depending on the convention chosen for the definition
of amplitudes. The value $\mathcal{C} = i/\sqrt{2}$ is consistent
with this work. The convention for the definition of the $b_{i}$ is
consistent with Ref. \cite{FTS92}. Inspection of Eqs.
(\ref{eq:MultExpF1}) to (\ref{eq:MultExpF4}) as well as the fact
that the function $\cos \theta$ is symmetric under the angular
reflection $\theta \rightarrow - \theta$ leads to the following
symmetry of the CGLN amplitudes
\begin{equation}
F_{i} \left(\theta\right) = F_{i} \left(- \theta\right) \mathrm{,}
\hspace*{2pt} i = 1,\ldots,4 \,.
\label{eq:CGLNAmpSymmetry}
\end{equation}
The combination of this symmetry property with the definitions of
transversity amplitudes (\ref{eq:b1BasicForm}) to
(\ref{eq:b4BasicForm}) deduces the following relations valid for the
$b_{i}$
\begin{equation}
b_{1} \left(\theta\right) = b_{2} \left(-\theta\right)\,, \quad
b_{3} \left(\theta\right) = b_{4} \left(-\theta\right) \,.
\label{eq:AmplitudeAngleRelation}
\end{equation}
It appears as if only two complex amplitudes are now necessary in
order to describe the photoproduction process, although this
achievement was obtained at the price of extending the angular
variable $\theta$ to unphysical values.

It should be noted that the equations relating transversity to CGLN amplitudes are linear, i.e.
\begin{equation}
 b_{i} = \sum \limits_{j = 1}^{4} \hat{T}_{ij} F_{j} \mathrm{.} \label{eq:CGLNTransversityEquivalent}
\end{equation}
This means that once a particular system of spin amplitudes is
known, the other one is as well.

For pseudoscalar meson photoproduction, there are 16 in principle
measurable polarization observables. These observables group into
the four classes of group S observables $\left\{ \sigma_{0}, \Sigma
, T , P \right\}$ containing also the unpolarized cross section
$\sigma_{0} = d \sigma/d \Omega$, beam-target (BT) observables
$\left\{E, F, G, H\right\}$, beam-recoil (BR) observables $\left\{
C_{x'}, C_{z'}, O_{x'}, O_{z'}\right\}$ and target-recoil (TR)
observables $\left\{T_{x'}, T_{z'}, L_{x'},
L_{z'}\right\}$~\cite{Barker75,Conventions}.

Table \ref{tab:chiangTrObs} summarizes the definitions of
observables used in this work. Since transversity amplitudes are
used in the following discussion, the observables are tabulated
exclusively in terms of the $b_{i}$. Independently of the system of
spin amplitudes used, every observable $\Omega$ is defined by a
profile function $\check{\Omega}$ that is a bilinear hermitian form
of the amplitudes. In order to obtain an observable from the
corresponding profile function, the latter has to be divided by the
unpolarized cross section. The conventions for observables used in
this work are consistent with those of Refs. \cite{Barker75} and
\cite{MAID}.

\begin{table}[ht]
\centering \caption{Polarization observables listed with sign
choices that are consistent with the MAID partial wave
analysis~\cite{Barker75,MAID}, for other conventions, see
Ref.~\cite{Conventions}. Observables are written using transversity
amplitudes.}
\begin{tabular}{cccccc}
\hline \hline Observable & Transversity representation & Type\\
\hline
$ I(\theta) = \sigma_{0}/\rho $ & $ \frac{1}{2} \left( \left| b_{1} \right|^{2} + \left| b_{2} \right|^{2} + \left| b_{3} \right|^{2} + \left| b_{4} \right|^{2} \right) $ &  \\
$ \check{\Sigma} $ & $ \frac{1}{2} \left( - \left| b_{1} \right|^{2} - \left| b_{2} \right|^{2} + \left| b_{3} \right|^{2} + \left| b_{4} \right|^{2} \right) $ & S \\
$ \check{T} $ & $ \frac{1}{2} \left( \left| b_{1} \right|^{2} - \left| b_{2} \right|^{2} - \left| b_{3} \right|^{2} + \left| b_{4} \right|^{2} \right) $ & \\
$ \check{P} $ & $ \frac{1}{2} \left( - \left| b_{1} \right|^{2} + \left| b_{2} \right|^{2} - \left| b_{3} \right|^{2} + \left| b_{4} \right|^{2} \right) $ & \\
\hline
$ \check{G} $ & $ \mathrm{Im} \left[ - b_{1} b_{3}^{\ast} - b_{2} b_{4}^{\ast} \right] $ & \\
$ \check{H} $ & $ - \mathrm{Re} \left[ b_{1} b_{3}^{\ast} - b_{2} b_{4}^{\ast} \right] $ & BT \\
$ \check{E} $ & $ - \mathrm{Re} \left[ b_{1} b_{3}^{\ast} + b_{2} b_{4}^{\ast} \right] $ & \\
$ \check{F} $ & $ \mathrm{Im} \left[ b_{1} b_{3}^{\ast} - b_{2} b_{4}^{\ast} \right] $ & \\
\hline
$ \check{O}_{x'} $ & $ - \mathrm{Re} \left[ - b_{1} b_{4}^{\ast} + b_{2} b_{3}^{\ast} \right] $ & \\
$ \check{O}_{z'} $ & $  \mathrm{Im} \left[ - b_{1} b_{4}^{\ast} - b_{2} b_{3}^{\ast} \right] $ & BR \\
$ \check{C}_{x'} $ & $  \mathrm{Im} \left[ b_{1} b_{4}^{\ast} - b_{2} b_{3}^{\ast} \right] $ & \\
$ \check{C}_{z'} $ & $  \mathrm{Re} \left[ b_{1} b_{4}^{\ast} + b_{2} b_{3}^{\ast} \right] $ & \\
\hline
$ \check{T}_{x'} $ & $ - \mathrm{Re} \left[ - b_{1} b_{2}^{\ast} + b_{3} b_{4}^{\ast} \right] $ & \\
$ \check{T}_{z'} $ & $ - \mathrm{Im} \left[ b_{1} b_{2}^{\ast} - b_{3} b_{4}^{\ast} \right] $ & TR \\
$ \check{L}_{x'} $ & $ - \mathrm{Im} \left[ - b_{1} b_{2}^{\ast} - b_{3} b_{4}^{\ast} \right] $ & \\
$ \check{L}_{z'} $ & $ \mathrm{Re} \left[ - b_{1} b_{2}^{\ast} - b_{3} b_{4}^{\ast} \right] $ & \\
\hline
\hline
\end{tabular}
\label{tab:chiangTrObs}
\end{table}

\section{Formalism for the study of ambiguities of the group S observables for a TPWA with $\ell \leq L$ } \label{sec:Ambiguities}

This section presents an ambiguity study of the group S observables.
The fundamental idea for this study, as presented in
Refs.~\cite{omel} and \cite{Gersten}, consists of exchanging the
angular variable $\cos \theta$ present in the multipole expansion of
Eqs.~(\ref{eq:MultExpF1}) to (\ref{eq:MultExpF4}) for $t =
\tan\theta/2$.

The fundamental trigonometric functions $\sin \theta$ and $\cos
\theta$ expressed in terms of $\tan\theta/2$ read~\cite{Gersten}
\begin{align}
\sin \theta &= \frac{2 \tan \frac{\theta}{2}}{ 1 + \tan^{2} \frac{\theta}{2}} \mathrm{,} \nonumber \\
\cos \theta &= \frac{1 - \tan^{2} \frac{\theta}{2}}{1 + \tan^{2} \frac{\theta}{2}} \mathrm{.} \label{eq:SinCosInTermsOfTan}
\end{align}
The relation for $\cos \theta$ can be formally inverted as follows
\begin{equation}
 \tan \frac{\theta}{2} = \left\{ \begin{array}{cl}  + \sqrt{\frac{1 - \cos \theta}{1 + \cos \theta}} \mathrm{,} & \theta \in \left[ 0, \hspace*{1pt} \pi \right] \\
  - \sqrt{\frac{1 - \cos \theta}{1 + \cos \theta}} \mathrm{,} & \theta \in \left[ - \pi, \hspace*{1pt} 0 \right] \end{array} \right. \mathrm{.} \label{eq:TanInTermsOfCos}
\end{equation}
Therefore $\cos \theta$ and $t = \tan\theta/2$ are recognized as
fully equivalent angular variables. As is shown in Ref.~\cite{omel}
and Appendix \ref{sec:AppendixA}, the multipole expansions of the
transversity amplitudes $b_{2}$ and $b_{4}$ up to a finite
truncation angular momentum $L$, take the form
\begin{align}
 b_{4} \left(\theta\right) &= \mathcal{C} \hspace*{1pt} \frac{\exp \left(i \frac{\theta}{2}\right)}{\left( 1 + t^{2} \right)^{L}} \hspace*{1pt} A_{2L}^{\prime} \left(t\right) \mathrm{,} \label{eq:b4Aprime} \\
 b_{2} \left(\theta\right) &= - \hspace*{1pt} \mathcal{C} \hspace*{1pt} \frac{\exp \left(i \frac{\theta}{2}\right)}{\left( 1 + t^{2} \right)^{L}} \hspace*{1pt} \left[ A_{2L}^{\prime} \left(t\right) + t D_{2L - 2}^{\prime} \left(t\right) \right] \mathrm{,} \label{eq:b2AprimeDprime}
\end{align}
when written in terms of $t$. $A_{2L}^{\prime} \left(t\right)$ and
$D_{2L-2}^{\prime} \left(t\right)$ are polynomials in $t$ with
generally complex coefficients. The definition of $B_{2L}^{\prime}
\left(t\right) = A_{2L}^{\prime} \left(t\right) + t D_{2L -
2}^{\prime} \left(t\right)$ simplifies
Eq.~(\ref{eq:b2AprimeDprime}). Once the amplitudes $b_{2}$ and
$b_{4}$ are known, the remaining functions $b_{1}$ and $b_{3}$ can
be obtained from Eq.~(\ref{eq:AmplitudeAngleRelation}). This fact
will be used repeatedly in the remaining discussion.
Appendix~\ref{sec:AppendixA} contains a derivation of the expression
for $A_{2L}^{\prime} \left( t \right)$ that reads
\begin{widetext}
\begin{align}
A_{2L}^{\prime} \left( t \right) &= \frac{1}{2} \sum_{\ell = 0}^{L} \Big\{ f_{\ell}^{(1)} (\ell+1) (\ell+2) (1 + t^{2})^{L-\ell} \,_{2}F_{1} \left( -\ell, -\ell-1; 2; - t^{2} \right) \nonumber \\
 & \quad \quad \quad \quad \hspace*{4pt} + f_{\ell}^{(2)} \ell (\ell-1) (1 + t^{2})^{L-\ell+2} \,_{2}F_{1} \left( -\ell+2, -\ell+1; 2; - t^{2} \right) \nonumber \\
 & \quad \quad \quad \quad \hspace*{4pt} + f_{\ell}^{(3)} \ell (\ell+1) (t + i)^{2} (1 + t^{2})^{L-\ell} \,_{2}F_{1} \left( -\ell+1, -\ell; 2; - t^{2} \right) \Big\} \mathrm{,} \label{eq:APrimeWithHypFunctions}
\end{align}
\end{widetext}
containing hypergeometric functions $\,_{2}F_{1} \left( a, b; c; Z
\right)$ (see also \cite{omel} and \cite{Gersten}).

$B_{2L}^{\prime} \left( t \right)$ composes by adding a similarly
looking expansion, i.e. $D_{2L-2}^{\prime} \left(t\right)$,
\begin{widetext}
\begin{align}
B_{2L}^{\prime} \left( t \right) &= A_{2L}^{\prime} \left(t\right) + \frac{t}{4} \sum_{\ell = 0}^{L} \Big\{ (if_{\ell}^{(4)}) \ell (\ell+1) (\ell+2) (\ell+3) (1 + t^{2})^{L-\ell} \,_{2}F_{1} \left( -\ell+1, -\ell-1; 3; - t^{2} \right) \nonumber \\
 & \quad \hspace*{1pt} + (if_{\ell}^{(5)}) (\ell-2) (\ell-1) \ell (\ell+1) (1 + t^{2})^{L-\ell+2} \,_{2}F_{1} \left( -\ell+3, -\ell+1; 3; - t^{2} \right) \nonumber \\ & \quad \hspace*{1pt} - (if_{\ell}^{(6)}) (\ell-1) \ell (\ell+1) (\ell+2) (t + i)^{2} (1 + t^{2})^{L-\ell} \,_{2}F_{1} \left( -\ell+2, -\ell; 3; - t^{2} \right) \Big\} \mathrm{,} \label{eq:BPrimeWithHypFunctions}
\end{align}
\end{widetext}
with the definitions of six partial wave coefficients (see
Appendix~\ref{sec:AppendixA}):
\begin{align}
f_{\ell}^{(1)} & = \ell M_{\ell+} + E_{\ell+} \mathrm{,} \label{eq:flNotation1MainText} \\
f_{\ell}^{(2)} &= (\ell+1) M_{\ell-} + E_{\ell-} \mathrm{,} \label{eq:f1Notation2MainText} \\
f_{\ell}^{(3)} &= (\ell+1) M_{\ell+} + \ell M_{\ell-} \mathrm{,} \label{eq:f1Notation3MainText} \\
f_{\ell}^{(4)} &= E_{\ell+} -  M_{\ell+} \mathrm{,} \label{eq:f1Notation4MainText} \\
f_{\ell}^{(5)} &= E_{\ell-} + M_{\ell-} \mathrm{,} \label{eq:flNotation5MainText} \\
f_{\ell}^{(6)} &= M_{\ell+} - E_{\ell+} - M_{\ell-} - E_{\ell-} \,.
\label{eq:flNotation6MainText}
\end{align}
Once the expressions (\ref{eq:APrimeWithHypFunctions}) and
(\ref{eq:BPrimeWithHypFunctions}) are evaluated for a specific $L$,
both reduce to polynomials in the variable $t$ having the finite
order $2L$ and complex coefficients $a_\ell, b_\ell$,
\begin{align}
A_{2L}^{\prime} \left(t\right) &= \sum_{\ell = 0}^{2L} a_{\ell} t^{\ell} \mathrm{,} \label{eq:APrimePolynomialform} \\
B_{2L}^{\prime} \left(t\right) &= \sum_{\ell = 0}^{2L} b_{\ell} t^{\ell} \mathrm{.} \label{eq:BPrimePolynomialform}
\end{align}
There appear $4L + 2$ expansion coefficients in
Eqs.~(\ref{eq:APrimePolynomialform}) and
(\ref{eq:BPrimePolynomialform}) that have to contain the same
information content as the $4L$ multipoles for a finite $L$ (see
Eqs.~(\ref{eq:MultExpF1}) to (\ref{eq:MultExpF4})). This counting
suggests that not all of the coefficients $a_{\ell}$ and $b_{\ell}$
are independent. This can be seen by first investigating
Eq.~(\ref{eq:b2AprimeDprime}) and noting that the polynomial
$D_{2L-2}^{\prime} \left(t\right)$ only has order $2L-2$, which
means that the leading coefficients of $A_{2L}^{\prime}
\left(t\right)$ and $B_{2L}^{\prime} \left(t\right)$ are equal (see
also (\ref{eq:BPrimeWithHypFunctions})). The term $t
D_{2L-2}^{\prime} \left(t\right)$ is zero for $t = 0$ and for every
order in $L$. Therefore also the free terms of $A_{2L}^{\prime}
\left(t\right)$ and $B_{2L}^{\prime} \left(t\right)$ are equal, i.e.
$A_{2L}^{\prime} \left(t = 0\right) \equiv B_{2L}^{\prime} \left(t =
0\right)$. Both facts are expressed in the relations
\begin{equation}
a_{2L} = b_{2L} \mathrm{,} \hspace*{8pt} a_{0} = b_{0} \mathrm{.} \label{eq:EqualityFreeAndLeadingTerms}
\end{equation}
A next convenient step is taken in Ref.~\cite{omel} by defining
normalized versions of $A_{2L}^{\prime} \left(t\right)$ and
$B_{2L}^{\prime} \left(t\right)$ by
\begin{align}
 A_{2L}^{\prime} \left(t\right) &= a_{2L} A_{2L} \left(t\right) \mathrm{,} \label{eq:A2LDefinition} \\
 B_{2L}^{\prime} \left(t\right) &= a_{2L} B_{2L} \left(t\right) \mathrm{,} \label{eq:B2LDefinition}
\end{align}
where the first identity $a_{2L} = b_{2L}$ of
Eq.~(\ref{eq:EqualityFreeAndLeadingTerms}) is already invoked. In
terms of the normalized polynomials $ A_{2L} \left(t\right)$ and $
B_{2L} \left(t\right)$ the amplitudes $b_{2}$ and $b_{4}$ take the
form
\begin{align}
 b_{4} \left(\theta\right) &= \mathcal{C} \hspace*{1pt} a_{2L} \hspace*{1pt} \frac{\exp \left(i \frac{\theta}{2}\right)}{\left( 1 + t^{2} \right)^{L}} \hspace*{1pt} A_{2L} \left(t\right) \mathrm{,} \label{eq:b4A} \\
 b_{2} \left(\theta\right) &= - \hspace*{1pt} \mathcal{C} \hspace*{1pt} a_{2L} \hspace*{1pt} \frac{\exp \left(i \frac{\theta}{2}\right)}{\left( 1 + t^{2} \right)^{L}} \hspace*{1pt} B_{2L} \left(t\right) \mathrm{,} \label{eq:b2B}
\end{align}
and both normalized polynomials can be written as
\begin{align}
A_{2L} \left(t\right) &= t^{2L} + \sum_{\ell=0}^{2L-1} \hat{a}_{\ell} t^{\ell} \mathrm{,} \label{eq:APolynomialform} \\
B_{2L} \left(t\right) &= t^{2L} + \sum_{\ell=0}^{2L-1} \hat{b}_{\ell} t^{\ell} \mathrm{.} \label{eq:BPolynomialform}
\end{align}
with new coefficients $\left\{\hat{a}_{\ell} = a_{\ell}/a_{2L} |
\ell = 0,\ldots,2L-1\right\}$ and $\left\{\hat{b}_{\ell} =
b_{\ell}/b_{2L} | \ell = 0,\ldots,2L-1\right\}$. The equality of the
free terms also survives for the normalized polynomials, i.e.
\begin{equation}
\hat{a}_{0} = \hat{b}_{0} \mathrm{.} \label{eq:CoeffEqualityHat}
\end{equation}
The number of independent complex coefficients in the present
formulation consisting of $a_{2L}$, $\hat{a}_{0}$ and
$\left\{\hat{a}_{\ell} | \ell \neq 0\right\}$ and
$\left\{\hat{b}_{\ell} | \ell \neq 0\right\}$ counts as $4L$ as it
should. It is now crucial to note~\cite{omel} that since $A_{2L}
\left(t\right)$ and $B_{2L} \left(t\right)$ are complex polynomials,
the fundamental theorem of algebra holds
and both decompose into products of their linear factors as follows
\begin{equation}
A_{2L} \left(t\right) = \prod_{k = 1}^{2L} \left( t - \alpha_{k} \right) \mathrm{,}
\hspace*{5pt} B_{2L} \left(t\right) = \prod_{k = 1}^{2L} \left( t - \beta_{k} \right) \mathrm{,}
\label{eq:LinearFactorDecomposition}
\end{equation}
with $ \left\{ \alpha_{k} \in \mathbb{C} | \hspace*{2pt} k =
1,\ldots,2L \right\} $ and $\left\{ \beta_{k} \in \mathbb{C} |
\hspace*{2pt} k = 1,\ldots,2L \right\}$ the complex roots of $A_{2L}
\left(t\right)$ and $B_{2L} \left(t\right)$, respectively. In terms
of a linear factorization (\ref{eq:LinearFactorDecomposition}), the
transversity amplitudes $b_{4}$ and $b_{2}$ become
\begin{align}
 b_{4} \left(\theta\right) &= \mathcal{C} \hspace*{1pt} a_{2L} \hspace*{1pt} \frac{\exp \left(i \frac{\theta}{2}\right)}{\left( 1 + t^{2} \right)^{L}} \hspace*{1pt} \prod_{k = 1}^{2L} \left( t - \alpha_{k} \right)  \mathrm{,} \label{eq:b4LinFact} \\
 b_{2} \left(\theta\right) &= - \hspace*{1pt} \mathcal{C} \hspace*{1pt} a_{2L} \hspace*{1pt} \frac{\exp \left(i \frac{\theta}{2}\right)}{\left( 1 + t^{2} \right)^{L}} \hspace*{1pt} \prod_{k = 1}^{2L} \left( t - \beta_{k} \right)  \mathrm{.} \label{eq:b2LinFact}
\end{align}
The equality of the free terms, i.e. $A_{2L} \left(t = 0\right)
\equiv B_{2L} \left(t = 0\right)$ yields (see
Eq.~(\ref{eq:LinearFactorDecomposition}))
\begin{equation}
\prod_{k=1}^{2L} \alpha_{k} = \prod_{k = 1}^{2L} \beta_{k} \mathrm{,} \label{eq:ConsistencyRelation}
\end{equation}
which will become an important relation in the following. Equation
(\ref{eq:ConsistencyRelation}) will be used to test if possible
ambiguities of the group S observables are consistent with the
underlying formalism. Therefore it is named the consistency
relation.

Another important object introduced in Ref.~\cite{omel} is the root
function $f(\theta,\alpha)$ defined by
\begin{align}
f \left( \theta \mathrm{,} \hspace*{1pt} \alpha \right) &= f
\left( \theta \mathrm{,} \hspace*{1pt} \alpha_{1},\ldots,\alpha_{2L} \right) \nonumber \\
&= \frac{\prod_{k = 1}^{2L} \left( \tan \frac{\theta}{2} -
\alpha_{k} \right)}{\left( 1 + \tan^{2} \frac{\theta}{2}
\right)^{L}} \mathrm{,} \label{eq:DeffOfThetaAlpha}
\end{align}
and $f(\theta,\beta) = f(\theta,\beta_{1},\ldots,\beta_{2L})$
accordingly. The following useful facts are valid for the root
function
\begin{align}
 f \left( \theta \mathrm{,} \hspace*{1pt} \alpha \right) |_{\theta=0} &= \prod_{k=1}^{2L} \alpha_{k} \mathrm{,} \label{eq:fFunctionAngularExtremes1} \\
 \lim_{\theta \rightarrow \pi}  f \left( \theta \mathrm{,} \hspace*{1pt} \alpha \right) &= 1 \mathrm{.} \label{eq:fFunctionAngularExtremes2}
\end{align}
When expressed using the root function, the amplitudes $b_{4}$ and $b_{2}$ acquire the simple form
\begin{align}
 b_{4} \left(\theta\right) &= \mathcal{C} \hspace*{1pt} a_{2L} \hspace*{1pt} \exp \left(i \frac{\theta}{2}\right) \hspace*{1pt} f \left( \theta \mathrm{,} \hspace*{1pt} \alpha \right)  \mathrm{,} \label{eq:b4f} \\
 b_{2} \left(\theta\right) &= - \hspace*{1pt} \mathcal{C} \hspace*{1pt} a_{2L} \hspace*{1pt} \exp \left(i \frac{\theta}{2}\right) \hspace*{1pt} f \left( \theta \mathrm{,} \hspace*{1pt} \beta \right)  \mathrm{.} \label{eq:b2f}
\end{align}
In order to obtain expressions for the remaining amplitudes $b_{3}$
and $b_{1}$, the angular reflection $\theta \rightarrow - \theta$ as
well as Eq.~(\ref{eq:AmplitudeAngleRelation}) have to be invoked.
Under reflection, the root functions behave as
\begin{align}
 f \left( - \theta \mathrm{,} \hspace*{1pt} \alpha \right) &= \frac{\prod_{k = 1}^{2L} \left( \tan\left(-\frac{\theta}{2}\right) - \alpha_{k} \right)}{\left( 1 + \tan^{2}\left(-\frac{\theta}{2}\right) \right)^{L}} \nonumber \\
&= \frac{\prod_{k = 1}^{2L} \left( - \tan \frac{\theta}{2} - \alpha_{k} \right)}{\left( 1 + \left(-\tan \frac{\theta}{2}\right)^{2} \right)^{L}} \nonumber \\
&= (-)^{2L} \frac{\prod_{k = 1}^{2L} \left( \tan \frac{\theta}{2} + \alpha_{k} \right)}{\left( 1 + \tan^{2} \frac{\theta}{2} \right)^{L}} \nonumber \\
&= f \left(  \theta \mathrm{,} \hspace*{1pt} - \alpha \right) \mathrm{.} \label{eq:fOfMinusTheta}
\end{align}
Therefore, the remaining transversity amplitudes can also be written in compact form as
\begin{align}
 b_{3} \left(\theta\right) &= b_{4} \left(-\theta\right) = \mathcal{C} \hspace*{1pt} a_{2L} \hspace*{1pt} \exp \left(- i \frac{\theta}{2}\right) \hspace*{1pt} f \left( \theta \mathrm{,} \hspace*{1pt} - \alpha \right)  \mathrm{,} \label{eq:b3f} \\
 b_{1} \left(\theta\right) &= b_{2} \left(-\theta\right) = - \hspace*{1pt} \mathcal{C} \hspace*{1pt} a_{2L} \hspace*{1pt} \exp \left(- i \frac{\theta}{2}\right) \hspace*{1pt} f \left( \theta \mathrm{,} \hspace*{1pt} - \beta \right)  \mathrm{.} \label{eq:b1f}
\end{align}
For the remaining discussion, it is important to consider the
behavior of the root functions under simultaneous complex
conjugation of all roots $\alpha \rightarrow \alpha^{\ast}$ or
$\beta \rightarrow \beta^{\ast}$
\begin{align}
 f \left( \theta \mathrm{,} \hspace*{1pt} \alpha^{\ast} \right) &= \frac{\prod_{k = 1}^{2L} \left( \tan \frac{\theta}{2} - \alpha_{k}^{\ast} \right)}{\left( 1 + \tan^{2} \frac{\theta}{2} \right)^{L}} \nonumber \\
&= \frac{\prod_{k = 1}^{2L} \left( \tan \frac{\theta}{2} - \alpha_{k} \right)^{\ast}}{\left[\left( 1 + \tan^{2} \frac{\theta}{2} \right)^{\ast}\right]^{L}} \nonumber \\
  &= \left[ \frac{\prod_{k = 1}^{2L} \left( \tan \frac{\theta}{2} - \alpha_{k} \right)}{\left( 1 + \tan^{2} \frac{\theta}{2} \right)^{L}}\right]^{\ast} \nonumber \\
&= f^{\ast} \left(  \theta \mathrm{,} \hspace*{1pt} \alpha \right) \mathrm{.} \label{eq:fOfConjugatedVariables}
\end{align}
Preceding the discussion of the ambiguity study of group S
observables, it is reasonable to compare the number of independent
real parameters in an ordinary truncated partial wave analysis and
the reformulated version. In an energy independent fit, the number
of independent real parameters for every order in $L$ counts as
\begin{equation}
 8 L - 1 \mathrm{,} \label{eq:IndepRealParameters}
\end{equation}
i.e. $4L$ complex multipoles with an undetermined overall phase.
There should be an equal number of parameters in the reformulated
version of the problem. The counting of the real degrees of freedom
represented by the roots $\{\alpha_{k}\}$ and $\{\beta_{k}\}$ gives
$8L$. Equation (\ref{eq:ConsistencyRelation}), reformulated as
follows
\begin{equation}
\prod_{k=1}^{2L} \alpha_{k} \left/ \prod_{k^{\prime} = 1}^{2L - 1} \beta_{k^{\prime}} \right. = \beta_{2L} \mathrm{,} \label{eq:ConsistencyRelationReformulated}
\end{equation}
reduces the number of independent real degrees of freedom of the
roots to $8L - 2$. There is one additional unknown complex variable
in the reformulation, $a_{2L}$. The modulus $\left|a_{2L}\right|$
can be determined from the forward scattering cross section $I(\pi)$
(see discussion below). The phase $\phi_{2L}$ of $a_{2L} =
\left|a_{2L}\right| e^{i \phi_{2L}}$ cannot be obtained by multipole
analysis. This leaves the anticipated number of $8L-1$ independent
real parameters for the reformulation of the multipole expansion.

What remains to be done before the ambiguities of the group S
observables are discussed is to establish a connection among the
complex coefficient $a_{2L}$ and the forward scattering cross
section $I(\pi)$. Utilizing the symmetry relation
(\ref{eq:AmplitudeAngleRelation}), the observable $I(\theta)$ takes
the form (see Table~\ref{tab:chiangTrObs})
\begin{align}
I \left(\theta\right) = \frac{1}{2} \Big( &\left| b_{2}
\left(-\theta\right) \right|^{2} + \left| b_{2} \left(\theta\right)
\right|^{2} + \left| b_{4} \left(-\theta\right) \right|^{2} + \left|
b_{4} \left(\theta\right) \right|^{2} \Big) \mathrm{.}
\label{eq:IntensityPMTheta}
\end{align}
In the limit $\theta \rightarrow \pi$, all root functions are unity
(see Eq.~(\ref{eq:fFunctionAngularExtremes2})). Therefore,
\begin{equation}
 I(\theta) |_{\theta \rightarrow \pi} = I(\pi) = 2 \left|
 \mathcal{C} \right|^{2} \left| a_{2L} \right|^{2} \,.
 \label{eq:DefIOfPi}
\end{equation}

In this work, the consistent value for $\mathcal{C}$ is $i/\sqrt{2}$
and Eq.~(\ref{eq:DefIOfPi}) yields $I(\pi) = \left| a_{2L}
\right|^{2}$. This is the anticipated relation connecting the
modulus $\left|a_{2L}\right|$ to the unpolarized cross section for
forward scattering.

With everything assembled until now, the possible ambiguities of
multipole solutions for the group S observables can be discussed.
Once the transversity amplitudes written in root functions (i.e.
Eqs.~(\ref{eq:b4f}), (\ref{eq:b2f}), (\ref{eq:b3f}) and
(\ref{eq:b1f})) are inserted into the group S observables of
Table~\ref{tab:chiangTrObs}, the latter take the form
\begin{widetext}
\begin{align}
 I \left(\theta\right) &= \frac{I \left(\pi\right)}{4} \left( \left| f \left( \theta \mathrm{,} \hspace*{1pt} - \beta \right) \right|^{2} + \left| f \left( \theta \mathrm{,} \hspace*{1pt} \beta \right) \right|^{2} + \left| f \left( \theta \mathrm{,} \hspace*{1pt} - \alpha \right) \right|^{2} + \left| f \left( \theta \mathrm{,} \hspace*{1pt} \alpha \right) \right|^{2} \right) \mathrm{,} \label{eq:IntInTermsOfF} \\
\check{\Sigma} \left(\theta\right) &= \frac{I \left(\pi\right)}{4} \left( - \left| f \left( \theta \mathrm{,} \hspace*{1pt} - \beta \right) \right|^{2} - \left| f \left( \theta \mathrm{,} \hspace*{1pt} \beta \right) \right|^{2} + \left| f \left( \theta \mathrm{,} \hspace*{1pt} - \alpha \right) \right|^{2} + \left| f \left( \theta \mathrm{,} \hspace*{1pt} \alpha \right) \right|^{2} \right) \mathrm{,} \label{eq:SigmaInTermsOfF} \\
\check{T} \left(\theta\right) &= \frac{I \left(\pi\right)}{4} \left( \left| f \left( \theta \mathrm{,} \hspace*{1pt} - \beta \right) \right|^{2} - \left| f \left( \theta \mathrm{,} \hspace*{1pt} \beta \right) \right|^{2} - \left| f \left( \theta \mathrm{,} \hspace*{1pt} - \alpha \right) \right|^{2} + \left| f \left( \theta \mathrm{,} \hspace*{1pt} \alpha \right) \right|^{2} \right) \mathrm{,} \label{eq:TInTermsOfF} \\
\check{P} \left(\theta\right) &= \frac{I \left(\pi\right)}{4} \left( - \left| f \left( \theta \mathrm{,} \hspace*{1pt} - \beta \right) \right|^{2} + \left| f \left( \theta \mathrm{,} \hspace*{1pt} \beta \right) \right|^{2} - \left| f \left( \theta \mathrm{,} \hspace*{1pt} - \alpha \right) \right|^{2} + \left| f \left( \theta \mathrm{,} \hspace*{1pt} \alpha \right) \right|^{2} \right) \mathrm{.} \label{eq:PInTermsOfF}
\end{align}
\end{widetext}
It can now be seen by inspection of the rule
(\ref{eq:fOfConjugatedVariables}) that the group S observables as
written above are invariant under the replacement
\begin{equation}
\alpha \rightarrow \alpha^{\ast}\,, \quad \beta \rightarrow
\beta^{\ast} \,,
\label{eq:DoubleAmbiguity}
\end{equation}
or, in more detail
\begin{equation}
\alpha_{i} \rightarrow \alpha_{i}^{\ast}\,,\quad  \beta_{j}
\rightarrow \beta_{j}^{\ast} \,, \quad i,j = 1,\ldots,2L \,.
\label{eq:DoubleAmbiguityDetail}
\end{equation}
In Ref.~\cite{omel}, this replacement rule was named the double
ambiguity. Once the newly obtained roots are resolved for the
multipoles, the new solution will generally be distinct from the
original one, but yield the same group S observables. Also, the new
solutions obtained via the double ambiguity transformation
automatically fulfill the consistency relation
(\ref{eq:ConsistencyRelation}). Complex conjugation of both sides of
Eq.~(\ref{eq:ConsistencyRelation}) yields
\begin{equation}
\prod_{k=1}^{2L} \alpha_{k}^{\ast} = \prod_{k = 1}^{2L} \beta_{k}^{\ast} \mathrm{,}  \label{eq:ConsistencyRelationConjugated}
\end{equation}
which proves the latter claim.

However, the double ambiguity is not the only possible ambiguity of
the group S observables, but every replacement similar to
Eq.~(\ref{eq:DoubleAmbiguityDetail}) with arbitrary subsets of
indices $\left\{i,j\right\}$ conjugated and all remaining indices
not conjugated leaves the group S observables invariant. The only
possibility to rule out those extra ambiguities is to check whether
or not they fulfill the consistency relation
(\ref{eq:ConsistencyRelation}). This fulfillment then would
correspond to a numerical accident and cannot be predicted. The
complex roots expressed in terms of phases read
\begin{equation}
\alpha_{k} = \left| \alpha_{k} \right| e^{i \varphi_{k}} \mathrm{,} \hspace*{10pt} \beta_{k} = \left| \beta_{k} \right| e^{i \psi_{k}} \mathrm{.} \label{eq:RootsPolarCoordinates}
\end{equation}
Using the quantities $\varphi_{k}$ and $\psi_{k}$, the fact that an
arbitrary combination of complex conjugations of the roots fulfills
the consistency relation (\ref{eq:ConsistencyRelation}) is
equivalent to the validity of the equation
\begin{equation}
 \pm \varphi_{1} \pm \ldots \pm \varphi_{2L} = \pm \psi_{1} \pm \ldots \pm \psi_{2L} \mathrm{,} \label{eq:ConsistencyRelationForPhases}
\end{equation}
for an arbitrary choice of sign combinations. The number of
candidates of additional solutions that can be formed by complex
conjugation of the roots $\{\alpha_{k}\}$ and $\{\beta_{k}\}$, since
$2^{2L}$ additional sets of $\{\alpha_{k}\}$ and $2^{2L}$ sets of
$\{\beta_{k}\}$ are possible, is $4^{2L}$. Therefore, the number of
$4^{2L}$ new potentially ambiguous solutions has to be tested
whether or not they fulfill the consistency relation
(\ref{eq:ConsistencyRelation}).

The sets of objects and formulas introduced until now facilitate an
ambiguity study of the group S observables. This procedure consists
of first beginning using a specific starting solution for multipoles
(for example taken from a partial wave analysis program) and then
computing the roots $\alpha$ and $\beta$. Once the roots are
calculated, additional sets of solutions are obtained by complex
conjugation, leaving the group S observables invariant. Next, for
all of these additional solutions, including the double ambiguity,
the behavior of the double polarization observables of the groups
BT, BR and TR under these new solutions has to be investigated. This
investigation should then yield a set of double polarization
observables that can remove all of the remaining ambiguities.
%
\section{Behavior of double polarization observables}
\label{sec:DoublePolarizationObs}
First, the behavior of the beam-target (BT) observables shall be
investigated. Inserting the transversity amplitude form of
Eqs.~(\ref{eq:b4f}), (\ref{eq:b2f}), (\ref{eq:b3f}) and
(\ref{eq:b1f}) into the definitions (Table~\ref{tab:chiangTrObs})
yields the expressions
\begin{widetext}
\begin{align}
\check{E} \left(\theta\right) &= - \frac{I\left(\pi\right)}{2} \hspace*{2pt} \mathrm{Re} \left[ - f \left(  \theta \mathrm{,} \hspace*{1pt} - \beta \right) f^{\ast} \left(  \theta \mathrm{,} \hspace*{1pt} - \alpha \right) - f \left(  \theta \mathrm{,} \hspace*{1pt}  \beta \right) f^{\ast} \left(  \theta \mathrm{,} \hspace*{1pt}  \alpha \right) \right] \mathrm{,} \label{eq:EReformulated} \\
\check{F} \left(\theta\right) &= \frac{I\left(\pi\right)}{2} \hspace*{2pt} \mathrm{Im} \left[ - f \left(  \theta \mathrm{,} \hspace*{1pt} - \beta \right) f^{\ast} \left(  \theta \mathrm{,} \hspace*{1pt} - \alpha \right) + f \left(  \theta \mathrm{,} \hspace*{1pt}  \beta \right) f^{\ast} \left(  \theta \mathrm{,} \hspace*{1pt}  \alpha \right) \right] \mathrm{,} \label{eq:FReformulated} \\
\check{G} \left(\theta\right) &= \frac{I\left(\pi\right)}{2} \hspace*{2pt} \mathrm{Im} \left[ f \left(  \theta \mathrm{,} \hspace*{1pt} - \beta \right) f^{\ast} \left(  \theta \mathrm{,} \hspace*{1pt} - \alpha \right) + f \left(  \theta \mathrm{,} \hspace*{1pt}  \beta \right) f^{\ast} \left(  \theta \mathrm{,} \hspace*{1pt}  \alpha \right) \right] \mathrm{,} \label{eq:GReformulated} \\
\check{H} \left(\theta\right) &= - \frac{I\left(\pi\right)}{2} \hspace*{2pt} \mathrm{Re} \left[ - f \left(  \theta \mathrm{,} \hspace*{1pt} - \beta \right) f^{\ast} \left(  \theta \mathrm{,} \hspace*{1pt} - \alpha \right) + f \left(  \theta \mathrm{,} \hspace*{1pt}  \beta \right) f^{\ast} \left(  \theta \mathrm{,} \hspace*{1pt}  \alpha \right) \right] \mathrm{.} \label{eq:HReformulated}
\end{align}
\end{widetext}
First of all it is important to note that the response of the BT
observables to the double ambiguity transformation
(\ref{eq:DoubleAmbiguity}) can be predicted. Consulting the rule
(\ref{eq:fOfConjugatedVariables}) describing the transformation of
the root functions under the double ambiguity, it is evident that
the observables $\check{F}$ as well as $\check{G}$, whose definition
involves the imaginary part, change sign in
Eqs.~(\ref{eq:FReformulated}) and (\ref{eq:GReformulated}). The
observables defined via real parts, i.e. $\check{E}$ and $\check{H}$
are invariant under the double ambiguity. Therefore they cannot
resolve it. For the angular boundary values $\theta = 0$ and $\pi$
the root functions behave as $f \left(\theta, \hspace*{1pt}
\alpha\right)|_{\theta=0} = \prod_{k} \alpha_{k}$ and $f
\left(\theta, \hspace*{1pt} \alpha\right)|_{\theta\rightarrow\pi} =
1$. Therefore, consulting Eqs.~(\ref{eq:EReformulated}) to
(\ref{eq:HReformulated}), the values taken by the BT observables on
the angular boundaries can be summarized, as is done in
Table~\ref{tab:BehaviourDPolObservables}.
%
%
%

Second, the beam-recoil (BR) observables
(Table~\ref{tab:chiangTrObs}) expressed by the root function $f$
read
\begin{widetext}
\begin{align}
\check{C}_{x^{\prime}} \left(\theta\right) &= \frac{I \left(\pi\right)}{2} \Big( \cos \theta \hspace*{2pt} \mathrm{Im} \left[ f \left(  \theta \mathrm{,} \hspace*{1pt} - \beta \right) f^{\ast} \left(  \theta \mathrm{,} \hspace*{1pt} \alpha \right) - f \left(  \theta \mathrm{,} \hspace*{1pt} \beta \right) f^{\ast} \left(  \theta \mathrm{,} \hspace*{1pt} - \alpha \right) \right] \nonumber \\
 & \quad \quad \quad \quad + \sin \theta \hspace*{2pt} \mathrm{Re} \left[ - f \left(  \theta \mathrm{,} \hspace*{1pt} \beta \right) f^{\ast} \left(  \theta \mathrm{,} \hspace*{1pt} - \alpha \right) - f \left(  \theta \mathrm{,} \hspace*{1pt} - \beta \right) f^{\ast} \left(  \theta \mathrm{,} \hspace*{1pt}  \alpha \right) \right] \Big) \mathrm{,} \label{eq:CxReformulated} \\
\check{C}_{z^{\prime}} \left(\theta\right) &= \frac{I \left(\pi\right)}{2} \Big( \cos \theta \hspace*{2pt} \mathrm{Re} \left[ f \left(  \theta \mathrm{,} \hspace*{1pt} - \beta \right) f^{\ast} \left(  \theta \mathrm{,} \hspace*{1pt} \alpha \right) + f \left(  \theta \mathrm{,} \hspace*{1pt} \beta \right) f^{\ast} \left(  \theta \mathrm{,} \hspace*{1pt} - \alpha \right) \right] \nonumber \\
 & \quad \quad \quad \quad + \sin \theta \hspace*{2pt} \mathrm{Im} \left[ f \left(  \theta \mathrm{,} \hspace*{1pt} - \beta \right) f^{\ast} \left(  \theta \mathrm{,} \hspace*{1pt}  \alpha \right) - f \left(  \theta \mathrm{,} \hspace*{1pt} \beta \right) f^{\ast} \left(  \theta \mathrm{,} \hspace*{1pt} - \alpha \right) \right] \Big) \mathrm{,} \label{eq:CzReformulated} \\
\check{O}_{x^{\prime}} \left(\theta\right) &= - \frac{I \left(\pi\right)}{2} \Big( \cos \theta \hspace*{2pt} \mathrm{Re} \left[ f \left(  \theta \mathrm{,} \hspace*{1pt} - \beta \right) f^{\ast} \left(  \theta \mathrm{,} \hspace*{1pt} \alpha \right) - f \left(  \theta \mathrm{,} \hspace*{1pt} \beta \right) f^{\ast} \left(  \theta \mathrm{,} \hspace*{1pt} - \alpha \right) \right] \nonumber \\
 & \quad \quad \quad \quad + \sin \theta \hspace*{2pt} \mathrm{Im} \left[ f \left(  \theta \mathrm{,} \hspace*{1pt} - \beta \right) f^{\ast} \left(  \theta \mathrm{,} \hspace*{1pt}  \alpha \right) + f \left(  \theta \mathrm{,} \hspace*{1pt} \beta \right) f^{\ast} \left(  \theta \mathrm{,} \hspace*{1pt} - \alpha \right) \right] \Big) \mathrm{,} \label{eq:OxReformulated} \\
\check{O}_{z^{\prime}} \left(\theta\right) &= - \frac{I \left(\pi\right)}{2} \Big( \cos \theta \hspace*{2pt} \mathrm{Im} \left[ f \left(  \theta \mathrm{,} \hspace*{1pt} - \beta \right) f^{\ast} \left(  \theta \mathrm{,} \hspace*{1pt} \alpha \right) + f \left(  \theta \mathrm{,} \hspace*{1pt} \beta \right) f^{\ast} \left(  \theta \mathrm{,} \hspace*{1pt} - \alpha \right) \right] \nonumber \\
 & \quad \quad \quad \quad + \sin \theta \hspace*{2pt} \mathrm{Re} \left[ f \left(  \theta \mathrm{,} \hspace*{1pt} \beta \right) f^{\ast} \left(  \theta \mathrm{,} \hspace*{1pt} - \alpha \right) - f \left(  \theta \mathrm{,} \hspace*{1pt} - \beta \right) f^{\ast} \left(  \theta \mathrm{,} \hspace*{1pt}  \alpha \right) \right] \Big) \mathrm{.} \label{eq:OzReformulated}
\end{align}
\end{widetext}
%
As all of them involve terms with real and imaginary parts, they all
change under the complex conjugation and, therefore, they all can
resolve the double ambiguity. Furthermore, the values of the
observables on the angular boundaries can be predicted. They are
listed in Table~\ref{tab:BehaviourDPolObservables}.


Finally, the target-recoil (TR) observables
(Table~\ref{tab:chiangTrObs}) are also expressed in terms of the
root function
\begin{widetext}
\begin{align}
\check{T}_{x^{\prime}} \left(\theta\right) &= - \frac{I \left(\pi\right)}{2} \Big( \cos \theta \hspace*{2pt} \mathrm{Re} \left[ f \left(  \theta \mathrm{,} \hspace*{1pt} - \beta \right) f^{\ast} \left(  \theta \mathrm{,} \hspace*{1pt} \beta \right) - f \left(  \theta \mathrm{,} \hspace*{1pt} - \alpha \right) f^{\ast} \left(  \theta \mathrm{,} \hspace*{1pt}  \alpha \right) \right] \nonumber \\
 & \quad \quad \quad \quad + \sin \theta \hspace*{2pt} \mathrm{Im} \left[ f \left(  \theta \mathrm{,} \hspace*{1pt} - \beta \right) f^{\ast} \left(  \theta \mathrm{,} \hspace*{1pt}  \beta \right) - f \left(  \theta \mathrm{,} \hspace*{1pt} - \alpha \right) f^{\ast} \left(  \theta \mathrm{,} \hspace*{1pt} \alpha \right) \right] \Big) \mathrm{,} \label{eq:TxReformulated} \\
\check{T}_{z^{\prime}} \left(\theta\right) &= \frac{I \left(\pi\right)}{2} \Big( \cos \theta \hspace*{2pt} \mathrm{Im} \left[ f \left(  \theta \mathrm{,} \hspace*{1pt} - \beta \right) f^{\ast} \left(  \theta \mathrm{,} \hspace*{1pt} \beta \right) - f \left(  \theta \mathrm{,} \hspace*{1pt} - \alpha \right) f^{\ast} \left(  \theta \mathrm{,} \hspace*{1pt}  \alpha \right) \right] \nonumber \\
 & \quad \quad \quad \quad + \sin \theta \hspace*{2pt} \mathrm{Re} \left[ - f \left(  \theta \mathrm{,} \hspace*{1pt} - \beta \right) f^{\ast} \left(  \theta \mathrm{,} \hspace*{1pt}  \beta \right) + f \left(  \theta \mathrm{,} \hspace*{1pt} - \alpha \right) f^{\ast} \left(  \theta \mathrm{,} \hspace*{1pt} \alpha \right) \right] \Big) \mathrm{,} \label{eq:TzReformulated} \\
\check{L}_{x^{\prime}} \left(\theta\right) &= \frac{I \left(\pi\right)}{2} \Big( \cos \theta \hspace*{2pt} \mathrm{Im} \left[ f \left(  \theta \mathrm{,} \hspace*{1pt} - \beta \right) f^{\ast} \left(  \theta \mathrm{,} \hspace*{1pt} \beta \right) + f \left(  \theta \mathrm{,} \hspace*{1pt} - \alpha \right) f^{\ast} \left(  \theta \mathrm{,} \hspace*{1pt}  \alpha \right) \right] \nonumber \\
 & \quad \quad \quad \quad + \sin \theta \hspace*{2pt} \mathrm{Re} \left[ - f \left(  \theta \mathrm{,} \hspace*{1pt} - \beta \right) f^{\ast} \left(  \theta \mathrm{,} \hspace*{1pt}  \beta \right) - f \left(  \theta \mathrm{,} \hspace*{1pt} - \alpha \right) f^{\ast} \left(  \theta \mathrm{,} \hspace*{1pt} \alpha \right) \right] \Big) \mathrm{,} \label{eq:LxReformulated} \\
\check{L}_{z^{\prime}} \left(\theta\right) &= \frac{I \left(\pi\right)}{2} \Big( \cos \theta \hspace*{2pt} \mathrm{Re} \left[ - f \left(  \theta \mathrm{,} \hspace*{1pt} - \beta \right) f^{\ast} \left(  \theta \mathrm{,} \hspace*{1pt} \beta \right) - f \left(  \theta \mathrm{,} \hspace*{1pt} - \alpha \right) f^{\ast} \left(  \theta \mathrm{,} \hspace*{1pt}  \alpha \right) \right] \nonumber \\
 & \quad \quad \quad \quad + \sin \theta \hspace*{2pt} \mathrm{Im} \left[ - f \left(  \theta \mathrm{,} \hspace*{1pt} - \beta \right) f^{\ast} \left(  \theta \mathrm{,} \hspace*{1pt}  \beta \right) - f \left(  \theta \mathrm{,} \hspace*{1pt} - \alpha \right) f^{\ast} \left(  \theta \mathrm{,} \hspace*{1pt} \alpha \right) \right] \Big) \mathrm{.} \label{eq:LzReformulated}
\end{align}
\end{widetext}
Again all of them change under the complex conjugation and are able
to resolve the double ambiguity. On the angular boundaries $\theta =
0$ and $\pi$ they take the values given in
Table~\ref{tab:BehaviourDPolObservables}.
%
%

%
\begin{widetext}
\begin{table}[h]
\centering
\caption{Angular boundary values of all double polarization observables.}
\begin{tabular}{c|c|c|c|c|c|c|c|c|c|c|c|c}
  & $\hspace*{2pt}$ $E$ $\hspace*{2pt}$ & $\hspace*{2pt}$ $F$ $\hspace*{2pt}$ & $\hspace*{2pt}$ $G$ $\hspace*{2pt}$ & $\hspace*{2pt}$ $H$ $\hspace*{2pt}$ & $\hspace*{2pt}$ $C_{x^{\prime}}$ $\hspace*{2pt}$ & $\hspace*{2pt}$ $C_{z^{\prime}}$ $\hspace*{2pt}$ & $\hspace*{2pt}$ $O_{x^{\prime}}$ $\hspace*{2pt}$ & $\hspace*{2pt}$ $O_{z^{\prime}}$ $\hspace*{2pt}$ & $\hspace*{2pt}$ $T_{x^{\prime}}$ $\hspace*{2pt}$ & $\hspace*{2pt}$ $T_{z^{\prime}}$ $\hspace*{2pt}$ & $\hspace*{2pt}$ $L_{x^{\prime}}$ $\hspace*{2pt}$ & $\hspace*{2pt}$ $L_{z^{\prime}}$ $\hspace*{2pt}$ \\
\hline
 $\theta = 0$ & $1$ & $0$ & $0$ & $0$ & $0$ & $+1$ & $0$ & $0$ & $0$  & $0$  & $0$ & $-1$ \\
\hline
 $\theta \rightarrow \pi$ & $1$ & $0$ & $0$ & $0$ & $0$ & $-1$ & $0$ & $0$ & $0$ & $0$ & $0$ & $+1$
\end{tabular}
\label{tab:BehaviourDPolObservables}
\end{table}
\end{widetext}

\section{A comparative numerical study for $L=1$}
\label{sec:Lequals1NumericalStudy}

\begin{widetext}
\begin{figure*}[ht]
\centering
        \includegraphics[width=0.435\textwidth]{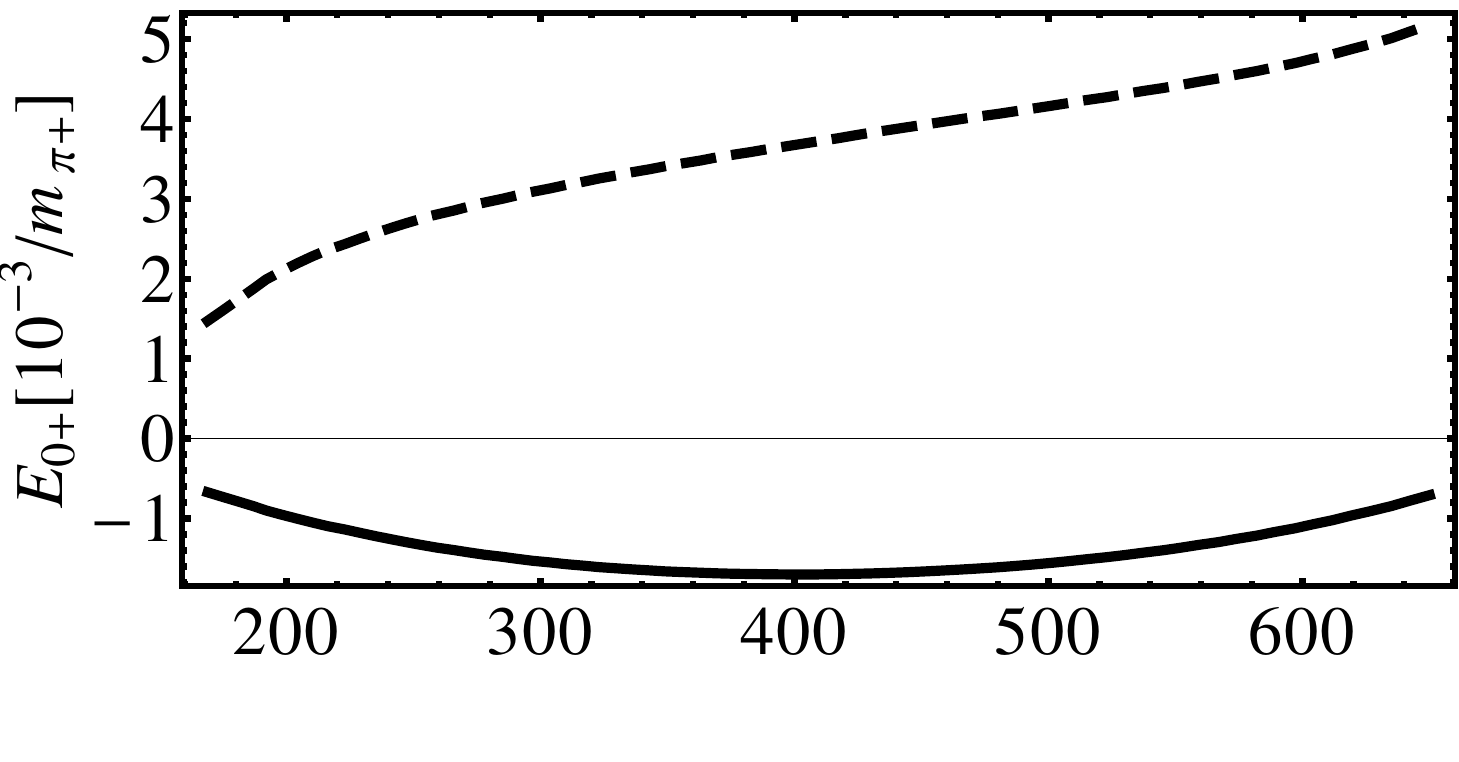}
        \hspace*{0.5cm}\includegraphics[width=0.435\textwidth]{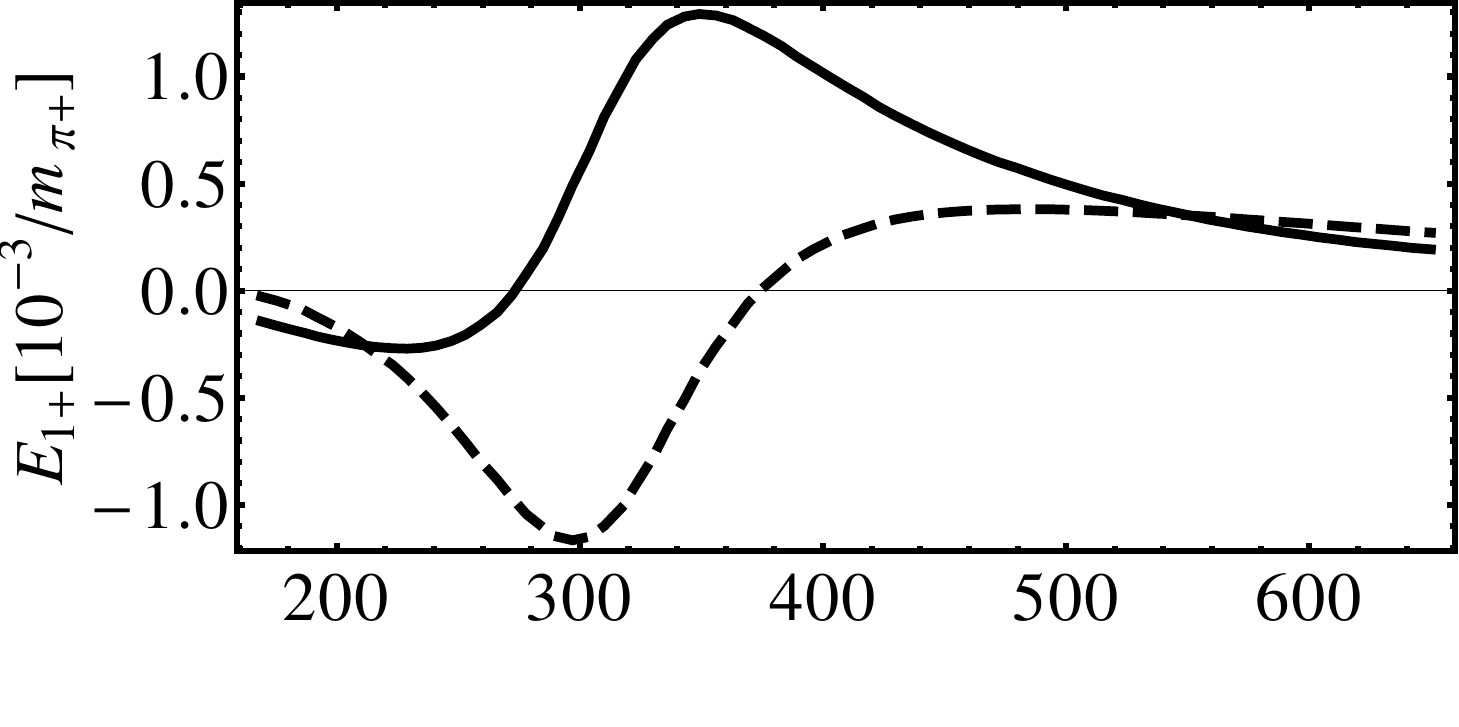}
        \includegraphics[width=0.435\textwidth]{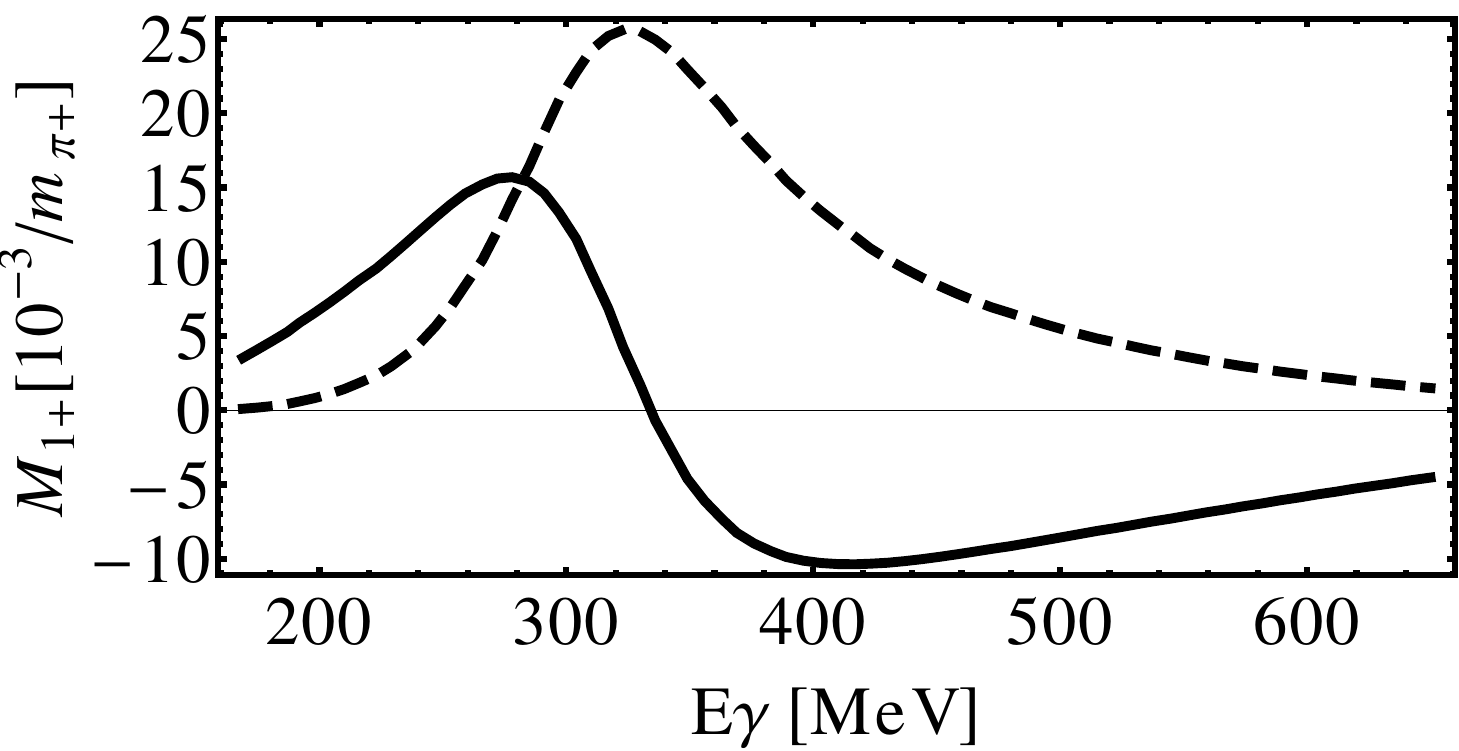}
        \hspace*{0.5cm}\includegraphics[width=0.435\textwidth]{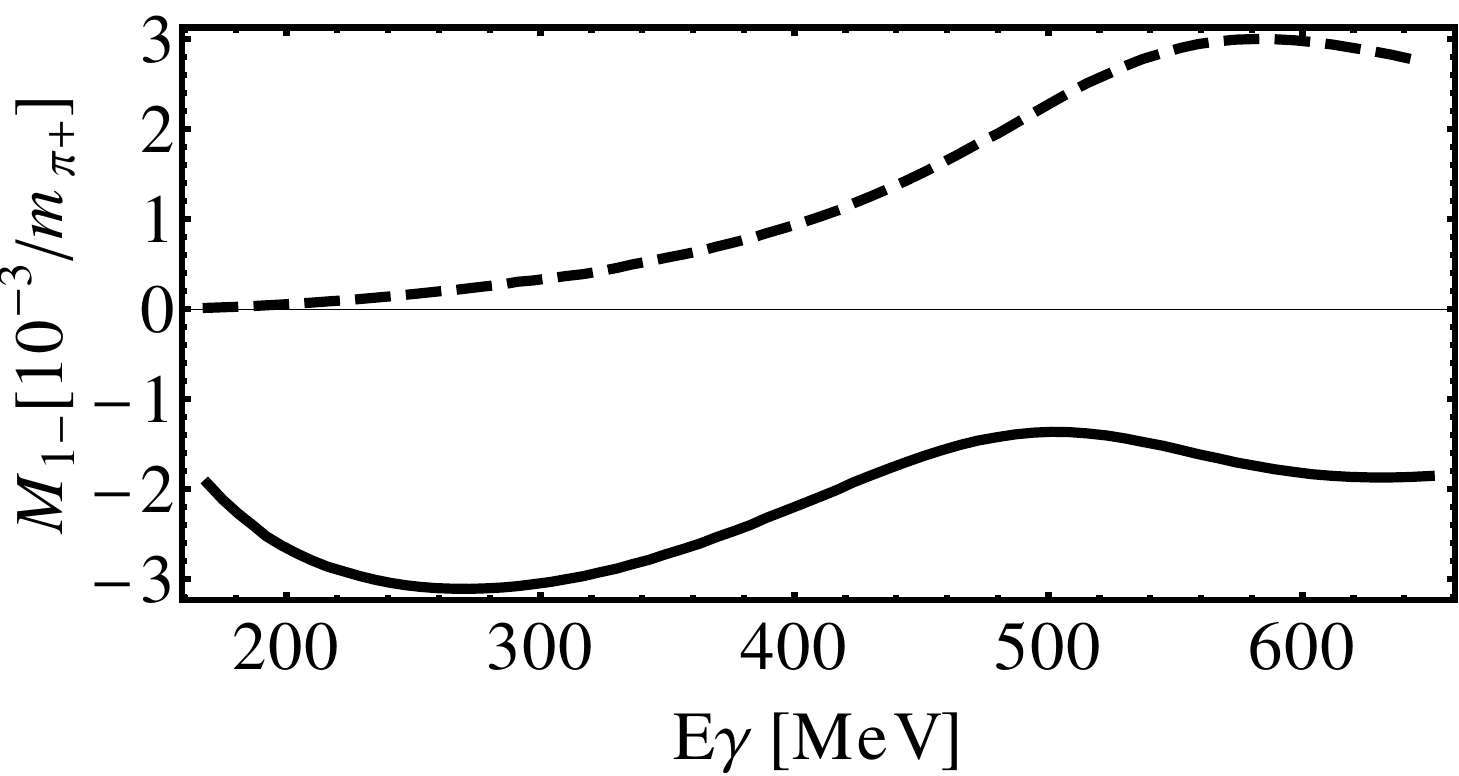}
\caption{Real (solid curves) and imaginary (dashed curves) parts of
the S- and P-wave multipoles of the MAID2007 solution. All
quantities are plotted versus the photon laboratory energy
$E_{\gamma}^{\mathrm{LAB}}$.} \label{fig:MAID2007StartingSolution}
\end{figure*}
\end{widetext}

This section contains the depiction of a numerical ambiguity study
performed using the formalism of Sec.~\ref{sec:Ambiguities} (see
\cite{omel} for a similar study). The case $L=1$ is considered. As
input for the study, multipoles are needed. The set of multipoles
used in this case originates from the MAID solution MAID2007 (see
\cite{MAID}), more precisely the channel $\gamma
p \rightarrow \pi^{0} p$. The multipoles corresponding to the S- and P-wave
approximation discussed here are
\begin{equation}
 \left\{ E_{0+}, \hspace*{2pt} E_{1+}, \hspace*{2pt} M_{1+}, \hspace*{2pt} M_{1-} \right\} \mathrm{.}
\end{equation}
For the starting MAID solution, the real and imaginary parts are
plotted in Fig.~\ref{fig:MAID2007StartingSolution}. The task now
consists of finding all possible sets of additional solutions that
leave the group S observables invariant and that are consistent with
the underlying formalism, i.e. fulfill the consistency relation
(\ref{eq:ConsistencyRelation}). The procedure starts with the MAID
solution. For $L = \ell_{\mathrm{max}} = 1$, i.e. S- and P-waves,
the normalized polynomials $A_{2L}\left(t\right)$ and
$B_{2L}\left(t\right)$ from Eqs.~(\ref{eq:A2LDefinition}) and
(\ref{eq:B2LDefinition}) become, with $t=\tan\theta/2$
\begin{align}
A_{2} \left( t \right) &= t^{2} + \hat{a}_{1} t + \hat{a}_{0} \nonumber \\
 &= t^{2} + 2 i \frac{2M_{1+} + M_{1-}}{E_{0+} - 3 E_{1+} - M_{1+} + M_{1-}} t \nonumber \\
 &\hspace*{9pt}+ \frac{E_{0+} + 3 E_{1+} + M_{1+} - M_{1-}}{E_{0+} - 3 E_{1+} - M_{1+} + M_{1-}} \mathrm{,} \label{eq:A2Lmax1} \\
B_{2} \left( t \right) &= t^{2} + \hat{b}_{1} t + \hat{b}_{0} \nonumber \\
 &= t^{2} + 2 i \frac{3 E_{1+} - M_{1+} + M_{1-}}{E_{0+} - 3 E_{1+} - M_{1+} + M_{1-}} t \nonumber \\
 &\hspace*{9pt}+ \frac{E_{0+} + 3 E_{1+} + M_{1+} - M_{1-}}{E_{0+} - 3 E_{1+} - M_{1+} + M_{1-}} \mathrm{.} \label{eq:B2Lmax1}
\end{align}
For this case the normalization coefficient is $a_{2} = b_{2} =
E_{0+} - 3 E_{1+} - M_{1+} + M_{1-}$. The modulus of the
normalization factor, or coefficient $a_{2}$ is given by
\begin{equation}
\left| a_{2} \right|^{2} = I(\pi) \,.
\label{eq:a2ModulusSquared}
\end{equation}
%
%

Therefore, as mentioned in Sec.~\ref{sec:Ambiguities}, in this
reformulation using polynomials, $a_{2}$ carries the undeterminable
overall phase of the multipoles. Once all coefficients, i.e.
$a_{2}$, $\hat{a}_{1}$, $\hat{a}_{0}$, $\hat{b}_{1}$ and
$\hat{b}_{0}$ are evaluated for each energy bin using the solution
MAID2007, the next step is to find the roots $\left\{ \alpha_{1},
\hspace*{1pt} \alpha_{2} \right\}$ for the polynomial
(\ref{eq:A2Lmax1}) and $\left\{ \beta_{1}, \hspace*{1pt} \beta_{2}
\right\}$ for (\ref{eq:B2Lmax1}). This task, as well as every other
numerical calculation mentioned in this section, was performed using
the computer algebra tool MATHEMATICA. The polynomials $A_{2}$ and
$B_{2}$ in this case acquire the linear factor decomposition
\begin{align}
A_{2} \left(t\right) &= \left( t - \alpha_{1} \right) \left( t - \alpha_{2} \right) \mathrm{,} \nonumber \\
B_{2} \left(t\right) &= \left( t - \beta_{1} \right) \left( t - \beta_{2} \right) \mathrm{.} \label{eq:LinearFactorDecompositionLmax1}
\end{align}
With the obtained roots it is easy to check that the consistency
relation~(\ref{eq:ConsistencyRelation}) for the case $L = 1$ reads
\begin{equation}
\alpha_{1} \alpha_{2} = \beta_{1} \beta_{2} \mathrm{,}
\label{eq:ConsistencyRelationLmax1}
\end{equation}
which is fulfilled for every energy bin by the starting MAID
solution. As mentioned in Sec.~\ref{sec:Ambiguities}, all candidates
for ambiguous solutions are constructed by complex conjugation of
roots. However, the argument in this section shall be made in an
equivalent way by using the phases of the roots~\cite{omel}. For the
latter, the consistency relation, defining $\alpha_{k} = \left|
\alpha_{k} \right| e^{i \varphi_{k}}$ and $\beta_{l} = \left|
\beta_{l} \right| e^{i \psi_{l}}$, reads
\begin{equation}
\varphi_{1} + \varphi_{2} = \psi_{1} + \psi_{2} \mathrm{.}
\label{eq:ConsistencyRelationLmax1Phases}
\end{equation}
The search for ambiguous solutions now consists of checking which
different choices of the signs in
Eq.~(\ref{eq:ConsistencyRelationLmax1Phases}) also yield a valid
equality. The arising possibilities can, for the case $L=1$, be
summarized by means of the equation
\begin{equation}
\pm \varphi_{1} \pm \varphi_{2} = \pm \psi_{1} \pm \psi_{2} \mathrm{.} \label{eq:AllCasesConsistencyRelationLmax1Phases}
\end{equation}
Before the above mentioned procedure is described further, it is
worth mentioning the way in which one can calculate the
corresponding multipoles, once new sets of phases and therefore also
roots are obtained. Phases and roots can yield the polynomial
coefficients. All that has to be done is to fully expand the linear
factorization~(\ref{eq:LinearFactorDecompositionLmax1}). The result,
relating roots and normalized polynomial coefficients, reads
\begin{align}
\hat{a}_{1} = - \alpha_{1} - \alpha_{2}\,,\quad \hat{a}_{0} = \alpha_{1} \alpha_{2} \,, \label{eq:RootsRelatedToCoefficientsA2} \\
\hat{b}_{1} = - \beta_{1} - \beta_{2}\,,\quad \hat{b}_{0} =
\beta_{1} \beta_{2} \,.
\label{eq:RootsRelatedToCoefficientsB2}
\end{align}
For the connection between coefficients and multipoles there exist
linear relations, as can be anticipated by inspection of
Eqs.~(\ref{eq:A2Lmax1}) and (\ref{eq:B2Lmax1}). For the case $L=1$
the following identities hold
\begin{widetext}
\begin{figure*}[ht]
\centering
\includegraphics[width=0.95\textwidth]{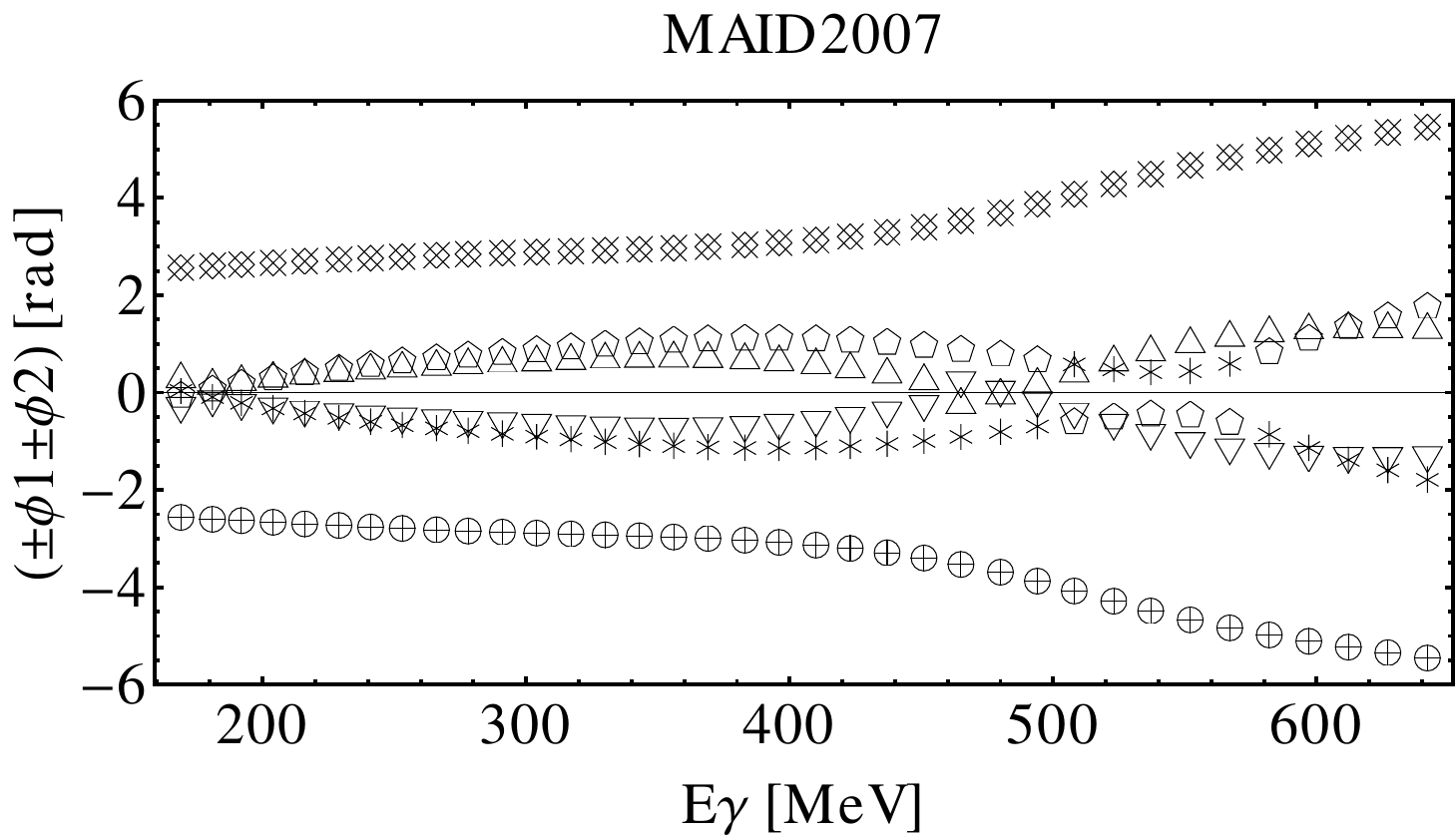}
\caption{Ambiguity diagram for the S- and P-wave multipoles (i.e. $L = \ell_{\mathrm{max}} = 1$) of
the MAID2007 solution as explained in the text. Plotted are different sign choices for linear
combinations of phases $\left\{\varphi_{1}, \hspace*{1pt} \varphi_{2}\right\}$ and
$\left\{\psi_{1}, \hspace*{1pt} \psi_{2}\right\}$, respectively. The scheme of labeling the
different linear combinations is the following:
  $\circ ( \varphi_{1} + \varphi_{2})$, $\bigtriangleup (\varphi_{1} - \varphi_{2})$,
  $\bigtriangledown (- \varphi_{1} + \varphi_{2})$, $\diamond (- \varphi_{1} - \varphi_{2})$,
 $+ (\psi_{1} + \psi_{2})$, ${\Large \ast} (\psi_{1} - \psi_{2})$, $\pentagon (- \psi_{1} + \psi_{2})$,
 $\times (- \psi_{1} -\psi_{2})$.} \label{fig:AmbiguityDiagramLmax1}
\end{figure*}
\end{widetext}
\begin{align}
E_{0+} &= \frac{1}{2} a_{2} \left( 1 + \hat{a}_{0} \right) \mathrm{,} \label{eq:E0plusInCoefficients} \\
E_{1+} &= \frac{1}{12} a_{2} \left( \hat{a}_{0} - 1 - i \hat{b}_{1} \right) \mathrm{,} \label{eq:E1plusInCoefficients} \\
M_{1+} &= \frac{1}{12} a_{2} \left( \hat{a}_{0} - 1 - 2 i \hat{a}_{1} + i \hat{b}_{1} \right) \mathrm{,} \label{eq:M1plusInCoefficients} \\
M_{1-} &= \frac{1}{6} a_{2} \left( 1 - \hat{a}_{0} - i \hat{a}_{1} - i \hat{b}_{1} \right) \mathrm{.} \label{eq:M1minusInCoefficients}
\end{align}
%
%
%

\begin{figure*}[h!]
\centering
\includegraphics[width=0.435\textwidth]{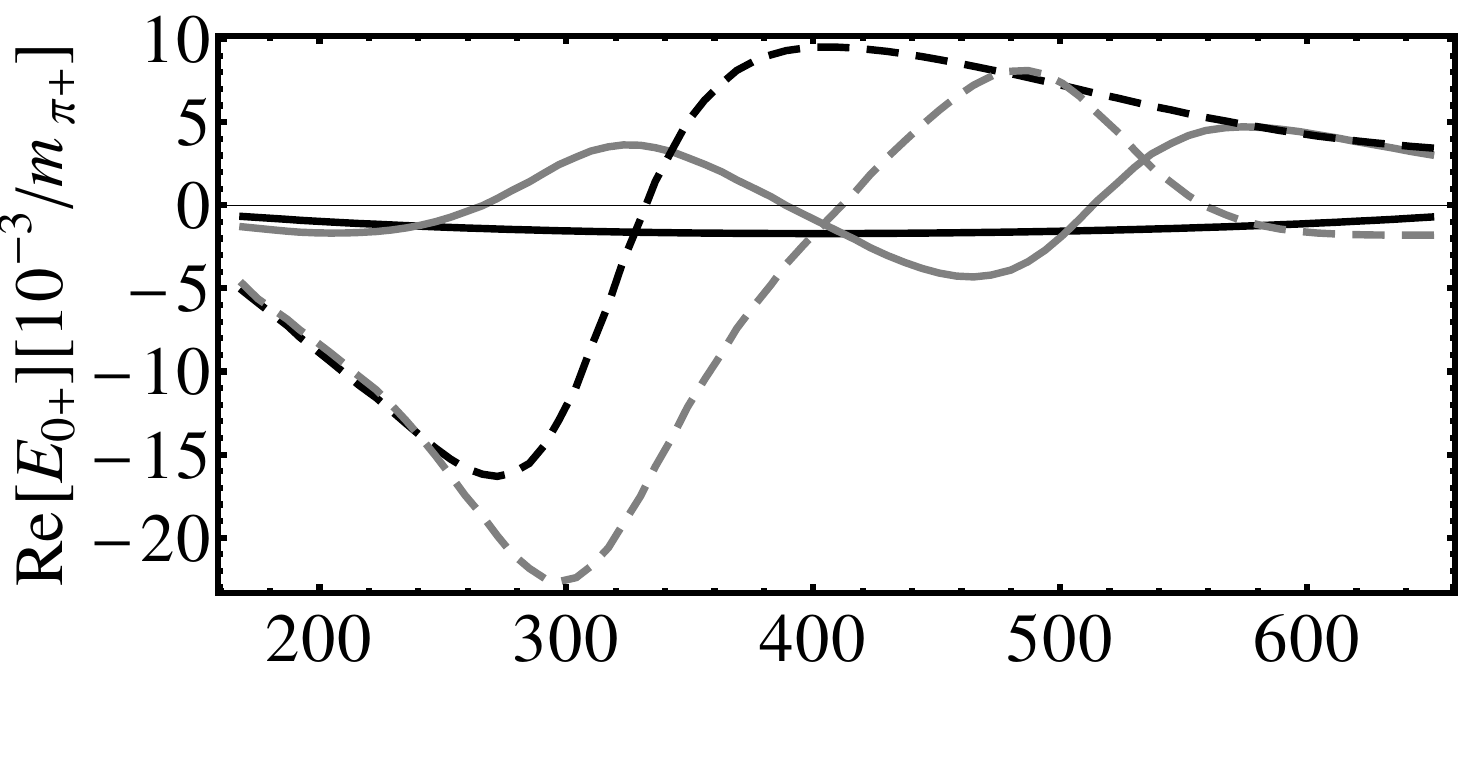}
\includegraphics[width=0.435\textwidth]{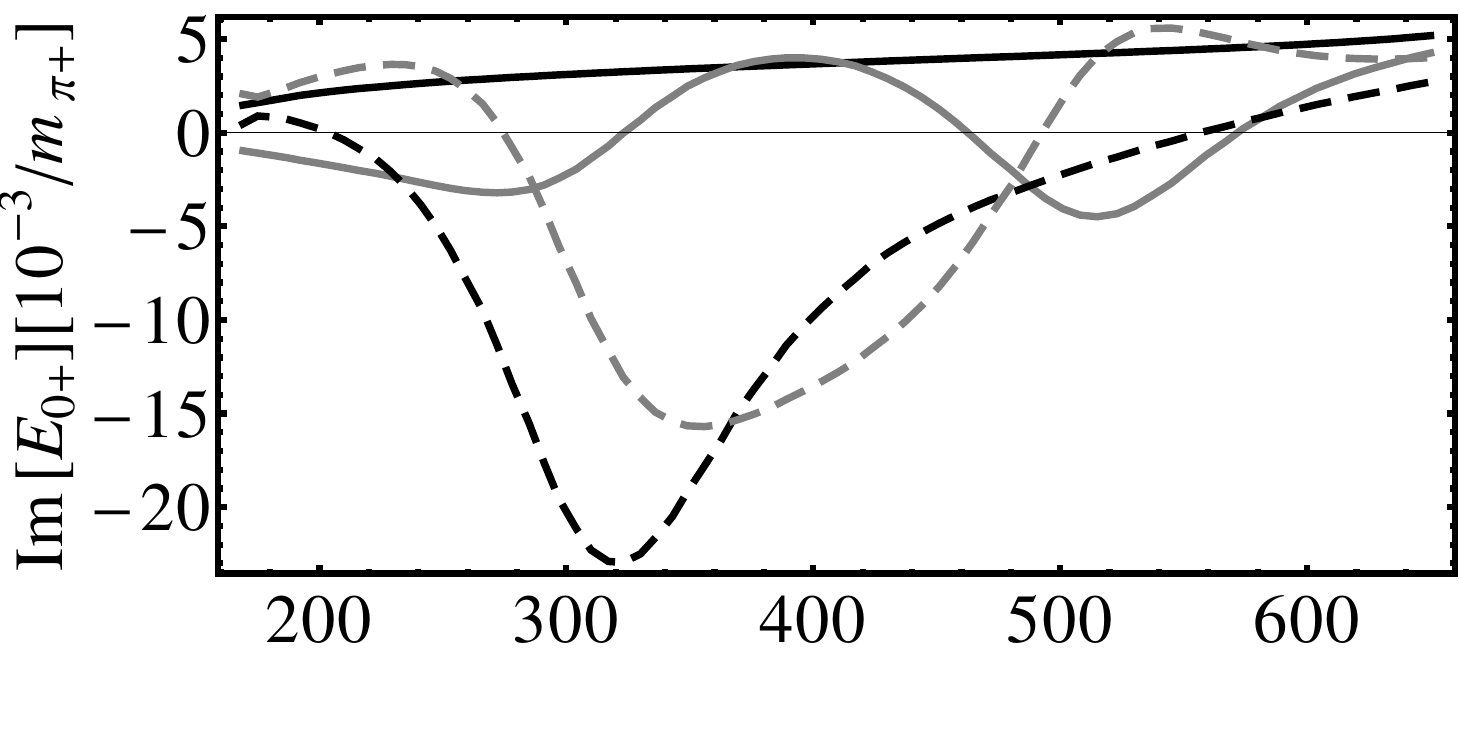}
\includegraphics[width=0.435\textwidth]{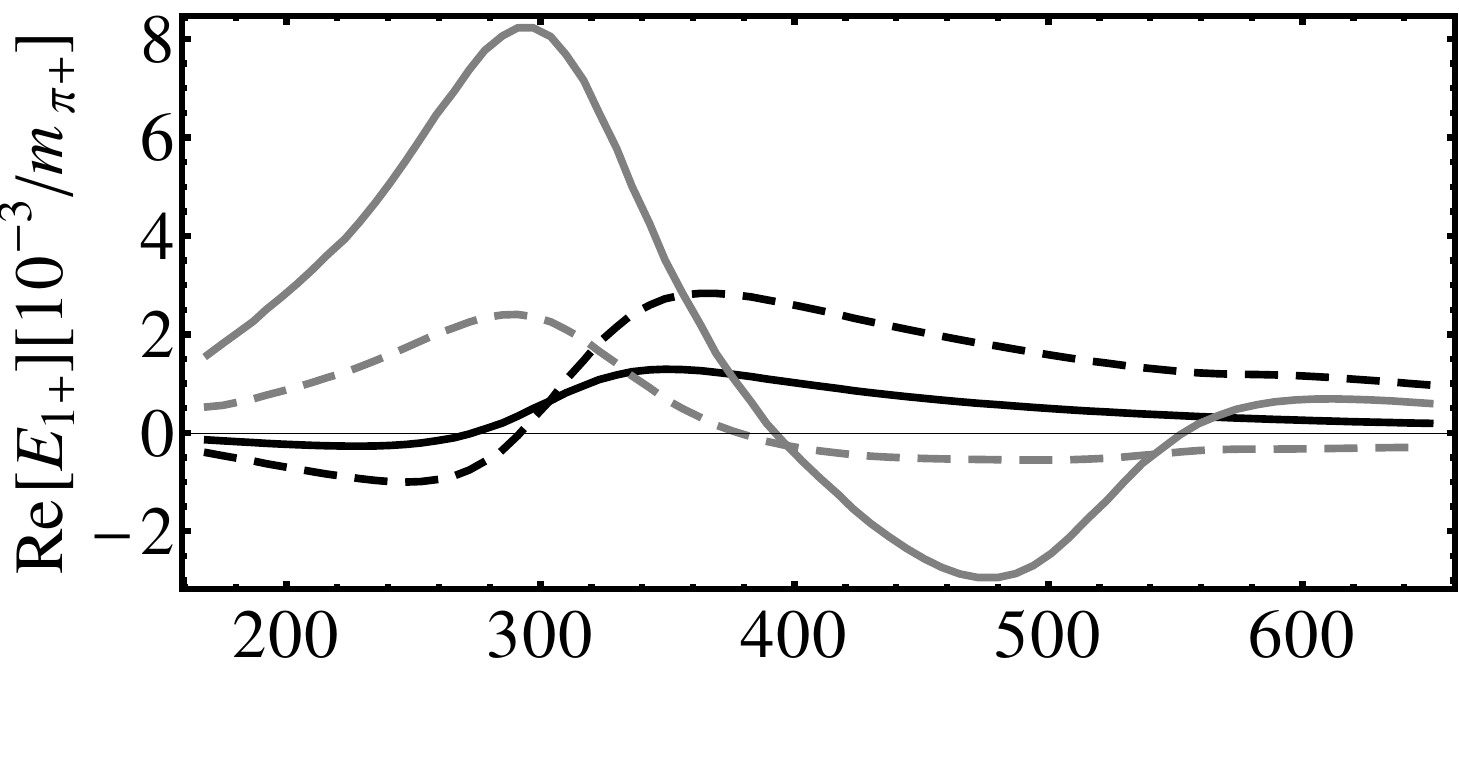}
\includegraphics[width=0.435\textwidth]{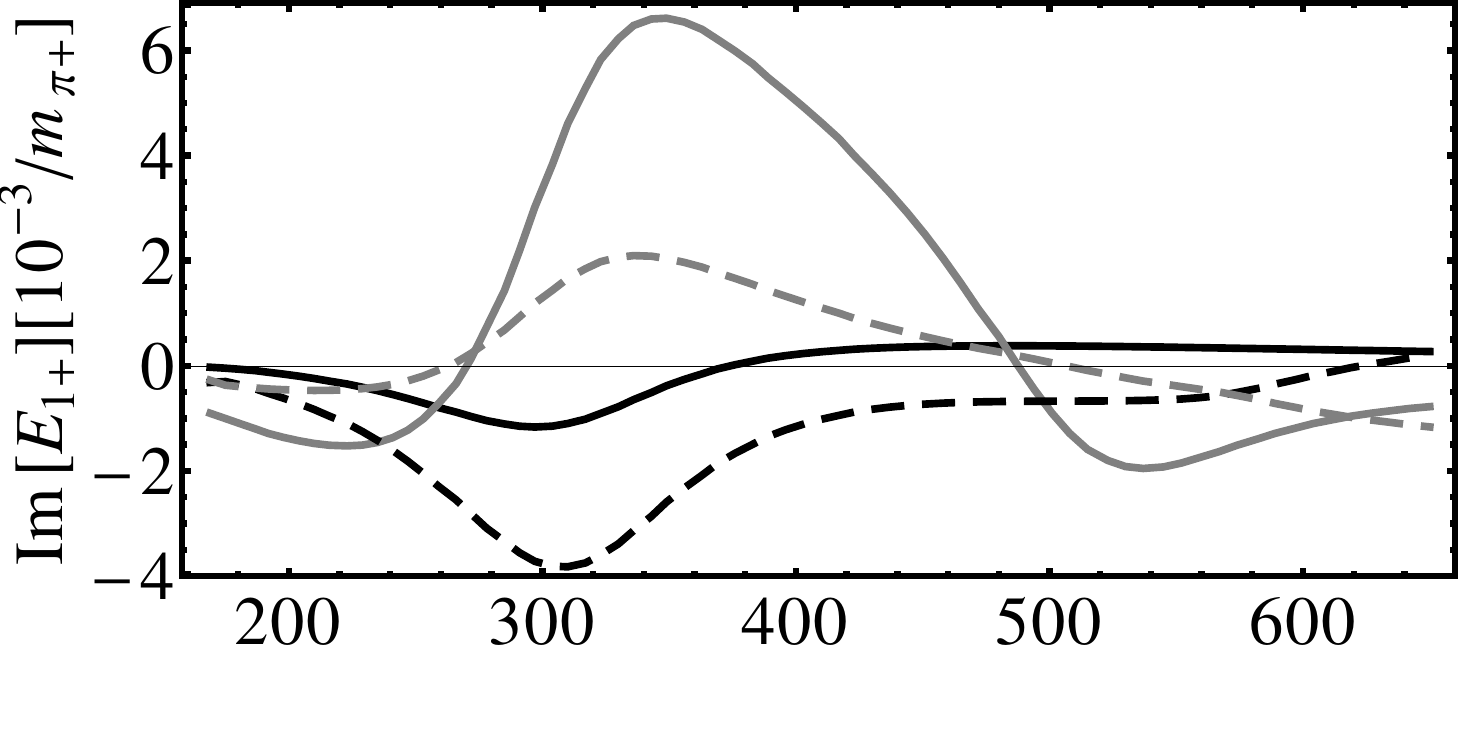}
\includegraphics[width=0.435\textwidth]{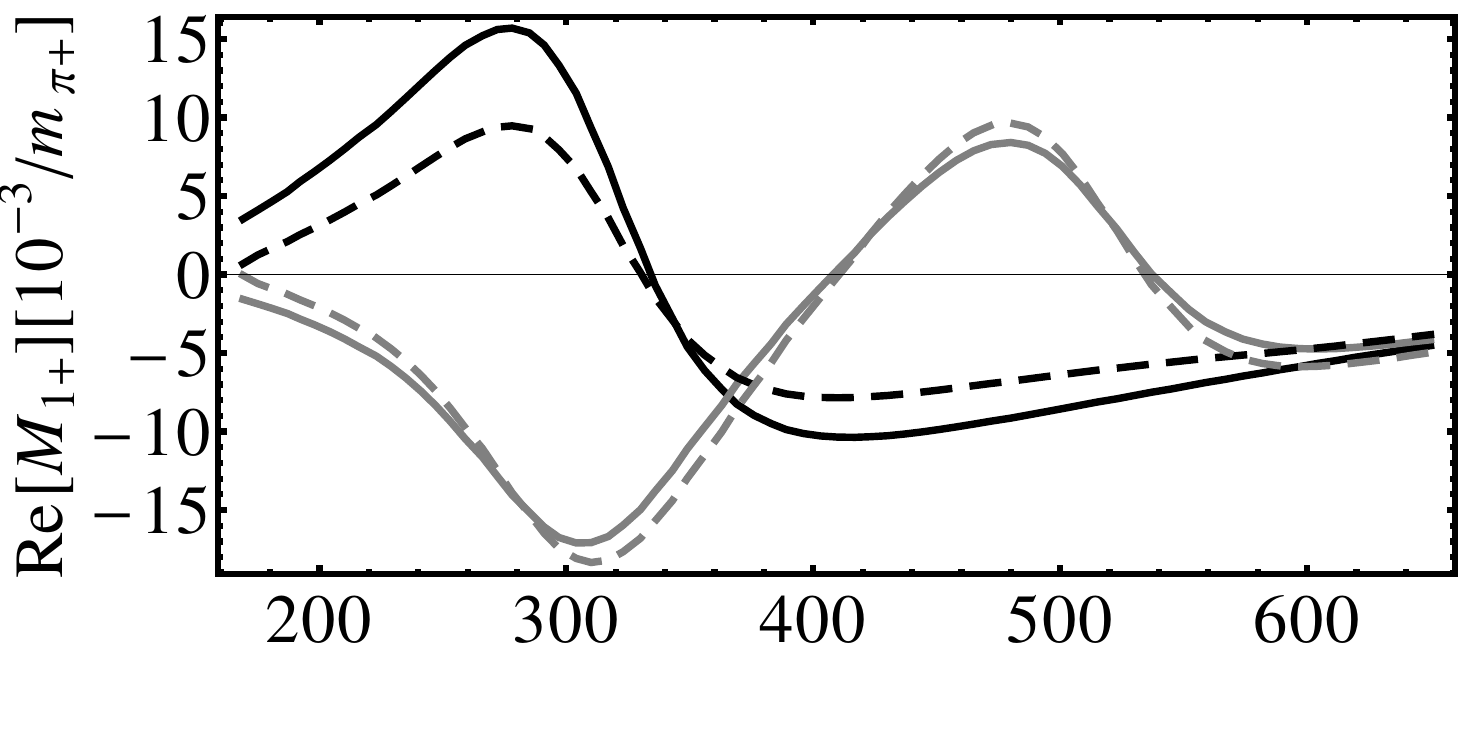}
\includegraphics[width=0.435\textwidth]{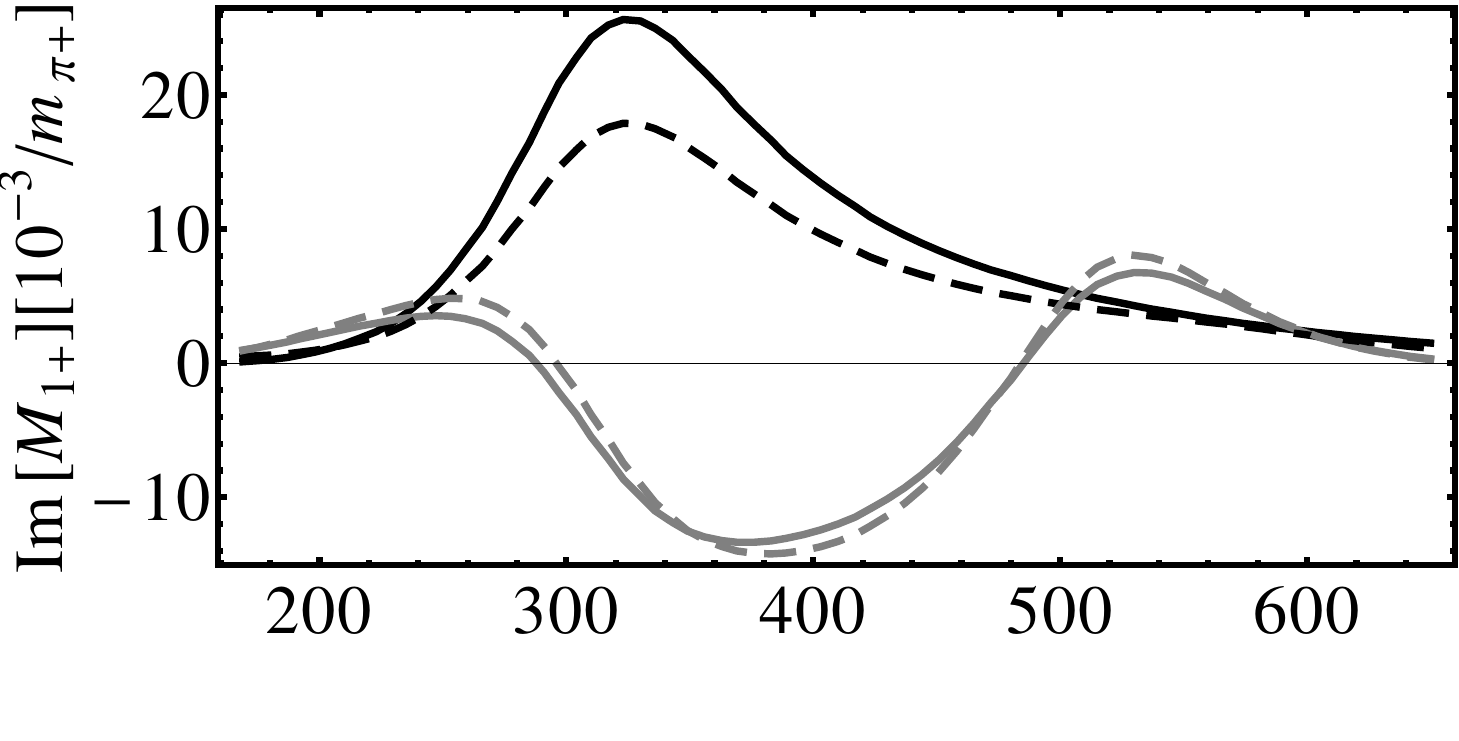}
\includegraphics[width=0.435\textwidth]{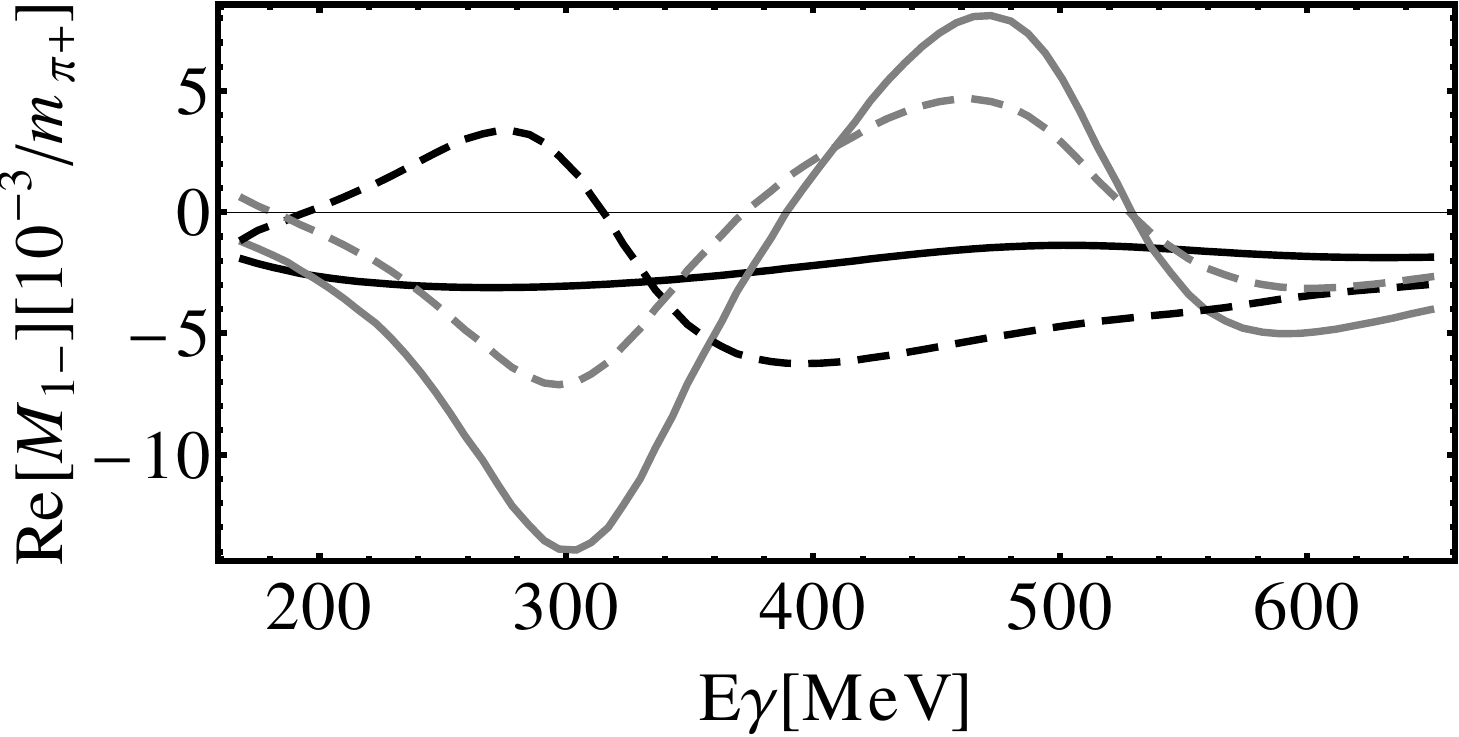}
\includegraphics[width=0.435\textwidth]{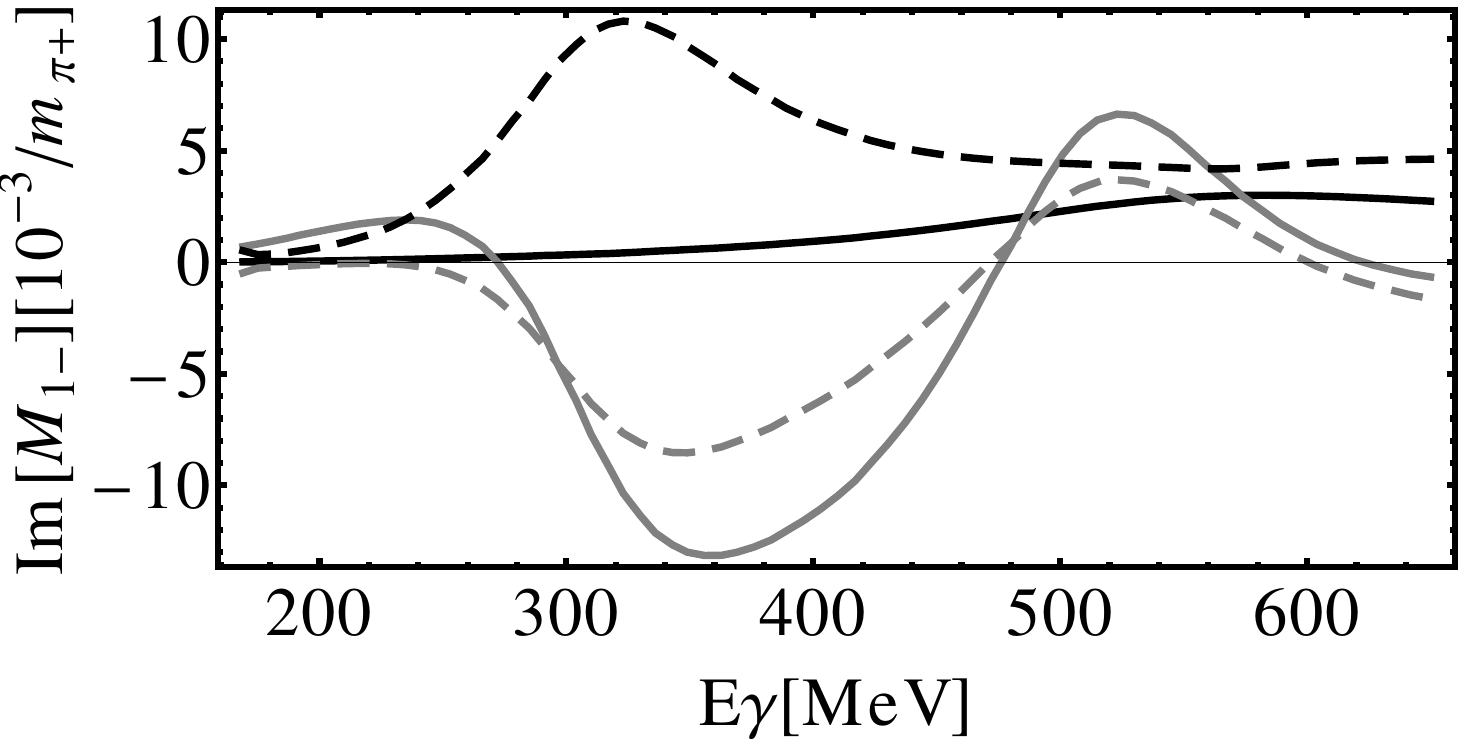}
\caption{S- and P-wave multipole ambiguities of the group S
observables extracted from Fig.~\ref{fig:AmbiguityDiagramLmax1}. The
starting solution is given by the solid black curves, the double
ambiguity by the solid grey curves. The accidental ambiguities due
to Eqs.~(\ref{eq:NumericalAccidentPhases1}) and
(\ref{eq:NumericalAccidentPhases2}) are plotted as dashed black and
dashed grey curves, respectively.}
\label{fig:MAID2007CompareMultipoles}
\end{figure*}

For $L=2$, Appendix~\ref{sec:AppendixC} contains the corresponding
relations as a more extensive example.
However, relations similar in structure to the examples in this
section can be derived for every finite order in $L$. Since roots
and multipoles are now established as fully equivalent sets of
complex variables, the description of the numerical ambiguity study
is continued. For each energy bin and for each combination of phases
appearing in Eq.~(\ref{eq:AllCasesConsistencyRelationLmax1Phases}),
the consistency relation has to be checked, separately.
The result of this procedure can be summarized by a plot that from
now on is referred to as the ambiguity diagram, given in
Fig.~\ref{fig:AmbiguityDiagramLmax1} (this type of diagram is also
given in Ref.~\cite{omel}). In this plot every possible case of sign
choices in the linear combinations of the phases $\left\{
\varphi_{1}, \hspace*{1pt} \varphi_{2} \right\}$ and $\left\{
\psi_{1}, \hspace*{1pt} \psi_{2} \right\}$ is drawn versus photon
laboratory energy $E_{\gamma}^{\mathrm{LAB}}$.
The caption of Fig.~\ref{fig:AmbiguityDiagramLmax1} provides the legend for the
symbols used in the ambiguity diagrams. Once a symbol representing
the left hand side of
Eq.~(\ref{eq:AllCasesConsistencyRelationLmax1Phases}) coincides with
one representing the right hand side, the consistency relation is
fulfilled and an ambiguity of the group S observables has to be
expected. For the starting solution this criterion is naturally
fulfilled for every energy bin, as depicted by the symbols
\begin{Large}$\circ$\end{Large} and $+$ in Fig.~\ref{fig:AmbiguityDiagramLmax1} (see
Eq.~(\ref{eq:ConsistencyRelationLmax1Phases})). Once all roots are
conjugated simultaneously, i.e.
\begin{equation}
\alpha \rightarrow \alpha^{\ast}\,,\quad \beta \rightarrow
\beta^{\ast} \,,
\label{eq:DoubleAmbiguityNumerical}
\end{equation}
the predicted double ambiguity is obtained (see
Sec.~\ref{sec:Ambiguities}). It corresponds to the symbols
$\diamond$ and $\times$ in Fig.~\ref{fig:AmbiguityDiagramLmax1}.
Additionally to the predictable ambiguities, numerically accidental
ambiguities are also possible. The remaining sign choices $(+,-)$
and $(-,+)$ are also given by their corresponding symbols in
Fig.~\ref{fig:AmbiguityDiagramLmax1}. As can be observed, symbols in
these two cases exactly coincide only for three cases at roughly
220, 515 and 615 MeV. Looking at the remaining energy bins, however,
it can be observed that the symbols are getting quite close.
Therefore, two additional ambiguous solutions can be expected for
the cases
\begin{equation}
\varphi_{1} - \varphi_{2} \approx - \psi_{1} + \psi_{2} \mathrm{,} \label{eq:NumericalAccidentPhases1}
\end{equation}
as well as
\begin{equation}
- \varphi_{1} + \varphi_{2} \approx \psi_{1} - \psi_{2} \mathrm{.} \label{eq:NumericalAccidentPhases2}
\end{equation}
%
Using Eqs.~(\ref{eq:RootsRelatedToCoefficientsA2}) to
(\ref{eq:M1minusInCoefficients}), the predicted as well as the
accidental ambiguities deduced from
Fig.~\ref{fig:AmbiguityDiagramLmax1} can be translated into
multipoles. The results are shown and explained in
Fig.~\ref{fig:MAID2007CompareMultipoles}.
%
%
%
\begin{figure*}[ht]
\centering
\includegraphics[width=0.435\textwidth]{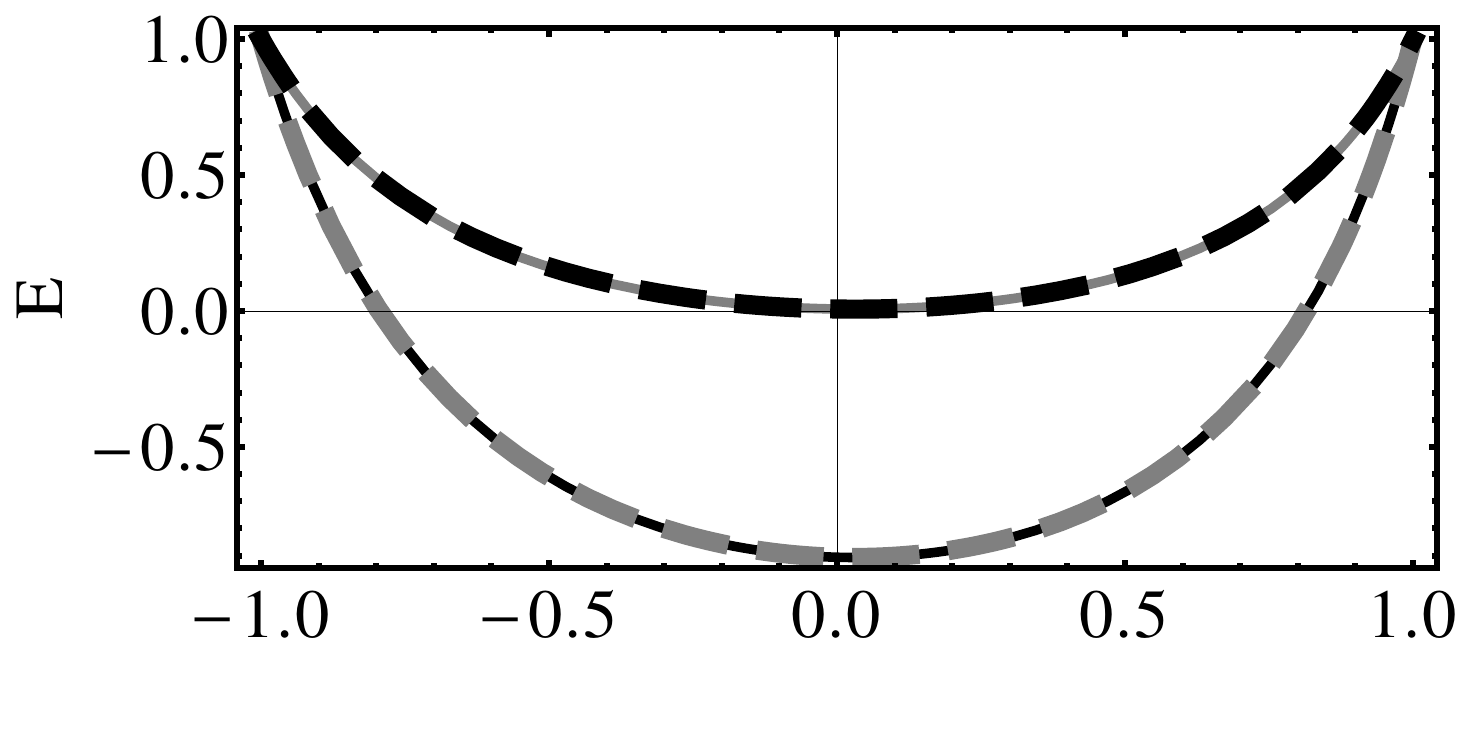}
\includegraphics[width=0.435\textwidth]{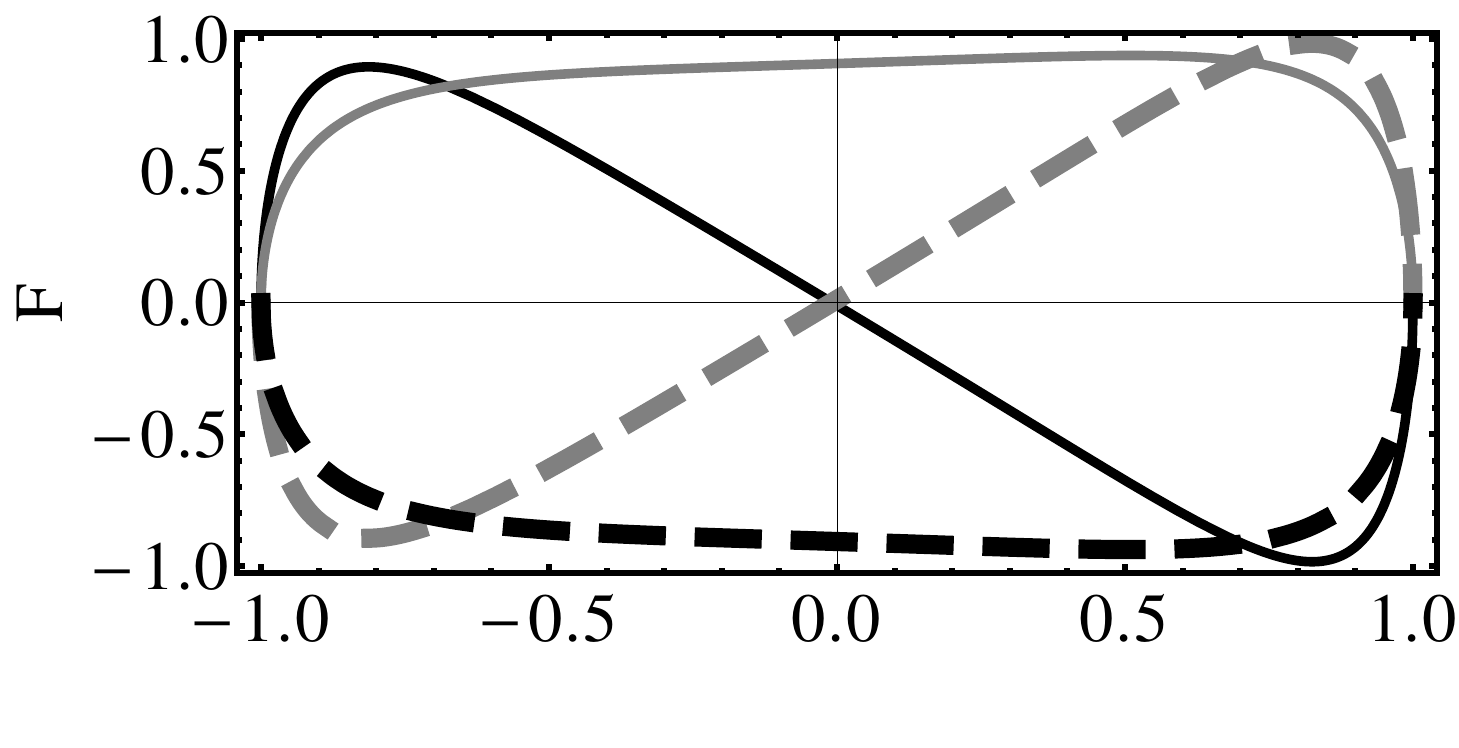}
\includegraphics[width=0.435\textwidth]{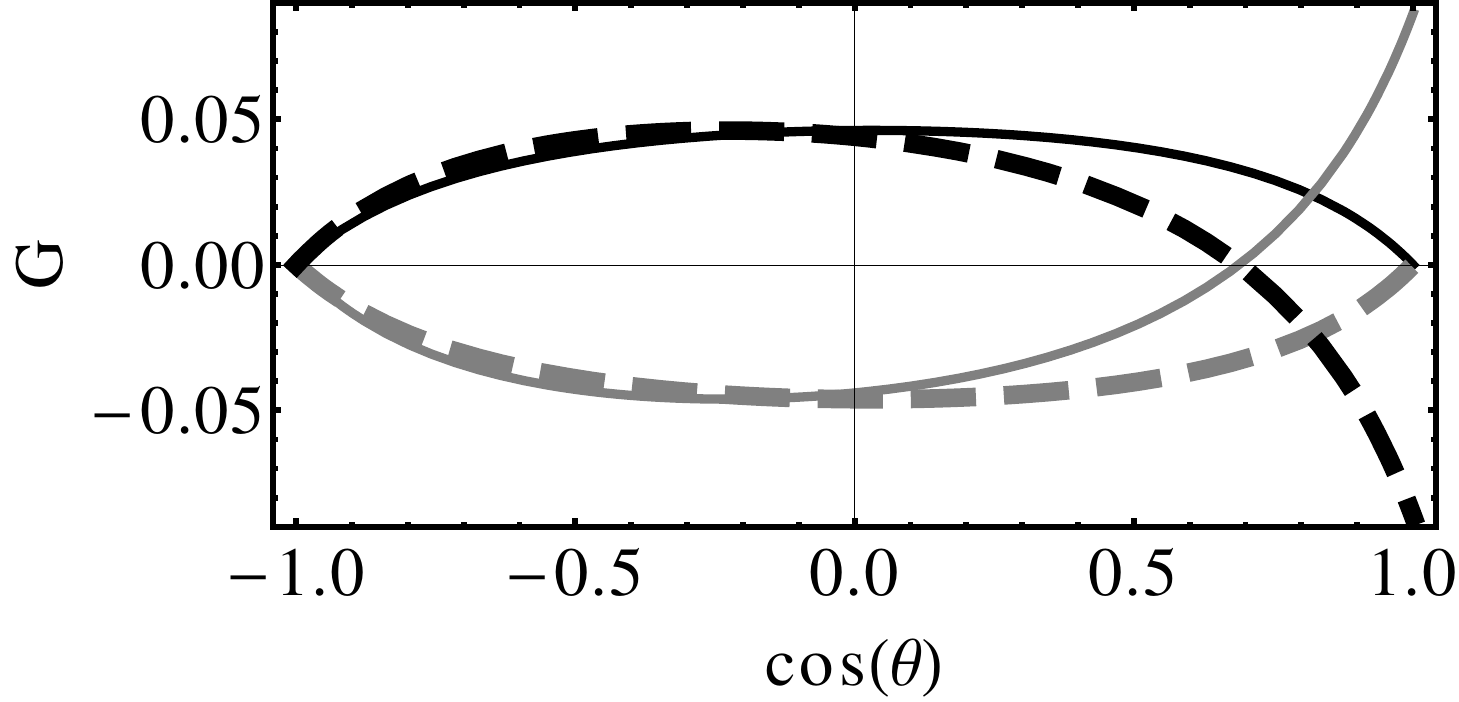}
\includegraphics[width=0.435\textwidth]{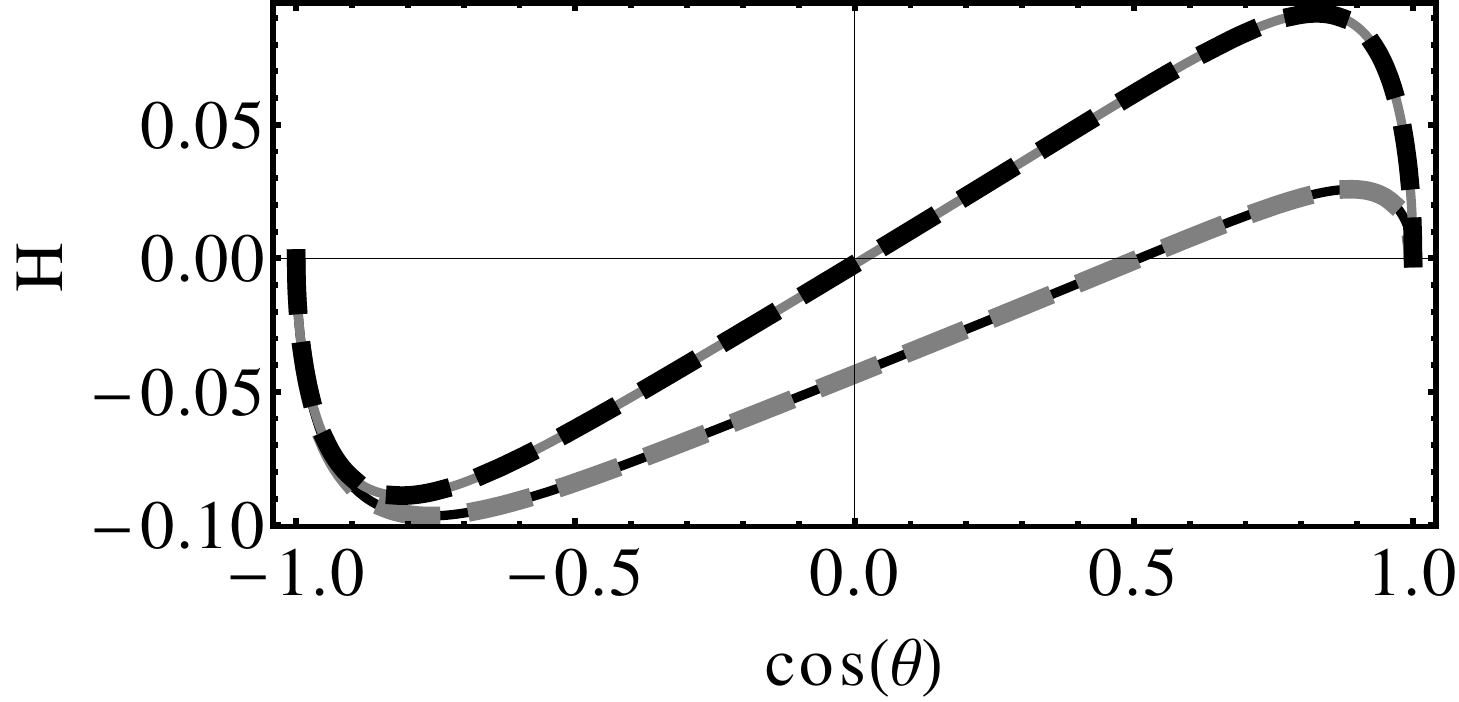}
\caption{Results of BT observables using the 4 different solutions
deduced from Fig.~\ref{fig:AmbiguityDiagramLmax1}. Therefore only S-
and P-wave multipoles contribute. The starting solution is given by
the solid black curves, the double ambiguity by the thick dashed
grey curves. The accidental ambiguities
(\ref{eq:NumericalAccidentPhases1}) and
(\ref{eq:NumericalAccidentPhases2}) are represented by the solid
grey and thick dashed black curves, respectively. For the
observables $F$ and $G$, all solutions are discriminable, which is
not true for $E$ and $H$. All observables are plotted versus the
angular variable $\cos \theta$. The energy bin of
$E_{\gamma}^{\mathrm{LAB}} = 253\,\mathrm{MeV}$ was chosen for this
picture.} \label{fig:MAID2007CompareObservables}
\end{figure*}
%
As can be observed, all solutions are smooth and distinct from each
other. Therefore, in case of a model independent truncated partial
wave analysis, the expectation is that for an S- and P-wave
truncation the group S observables will not be able to distinguish
among the four solutions plotted in
Fig.~\ref{fig:MAID2007CompareMultipoles}. Once
Eqs.~(\ref{eq:IntInTermsOfF}) to (\ref{eq:PInTermsOfF}) are used to
calculate group S observables, it can be seen that the results for
the four different solutions exactly coincide (this can also be seen
from the formalism of Sec.~\ref{sec:Ambiguities}).
%
%
The ingredient that is needed in order to decide which of the four
solution candidates is the correct one are double polarization
observables. Since the observables of the class BT are the most
experimentally accessible ones, the focus is drawn to them. Fig.
\ref{fig:MAID2007CompareObservables} shows plots that result from
the application of Eqs.~(\ref{eq:EReformulated}) to
(\ref{eq:HReformulated}) to the four ambiguous solutions deduced in
this study. The BT observables are calculated and drawn such that
they can be graphically distinguished from each other. The energy
bin $E_{\gamma}^{\mathrm{LAB}} = 253 \, \mathrm{MeV}$ was chosen as
an example. As can be observed, for the observables $E$ and $H$, the
starting solution and the double ambiguity as well as both
accidental ambiguities exactly coincide. Therefore it is expected
that in a truncated partial wave analysis, data for both observables
will not be able to distinguish among the corresponding ambiguities,
in particular not between the double ambiguity and the starting solution.
$F$ and $G$ on the other hand show differing curves for all four solutions, which means
that both observables should be capable of yielding the correct
unique solution in the performed fit. Another feature that can be
observed for the observable $G$ is that both solutions corresponding
to the accidental ambiguities postulated in this section show a
behavior that contradicts the rules deduced in
Sec.~\ref{sec:DoublePolarizationObs}, i.e. $G$ does not approach $0$
for $\cos \theta \rightarrow 1$. Inspecting the ambiguity diagram
for $E_{\gamma}^{\mathrm{LAB}} = 253 \, \mathrm{MeV}$, the phases
are close but do not completely overlap and the consistency relation
is not exactly fulfilled. With high precision data this can be
distinguished, for data with sizeable errors it could well show up
as an additional ambiguity.

As a result of the ambiguity study presented until now, it should be
stated that in the context of a truncated partial wave analysis with
$L = 1$, i.e. S- and P-waves,
\begin{figure}[ht]
\centering
\includegraphics[width=0.4775\textwidth]{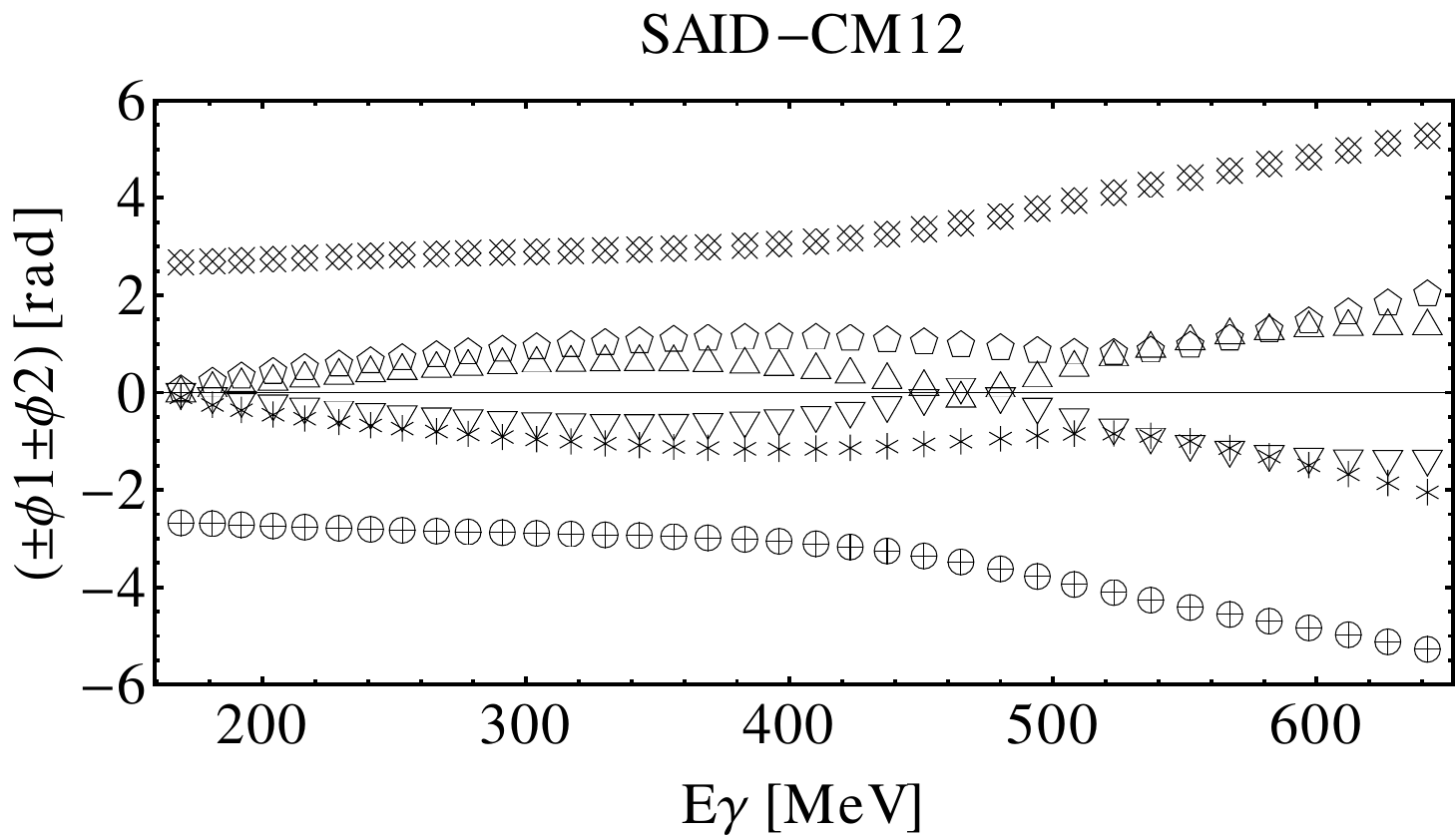}
\includegraphics[width=0.4775\textwidth]{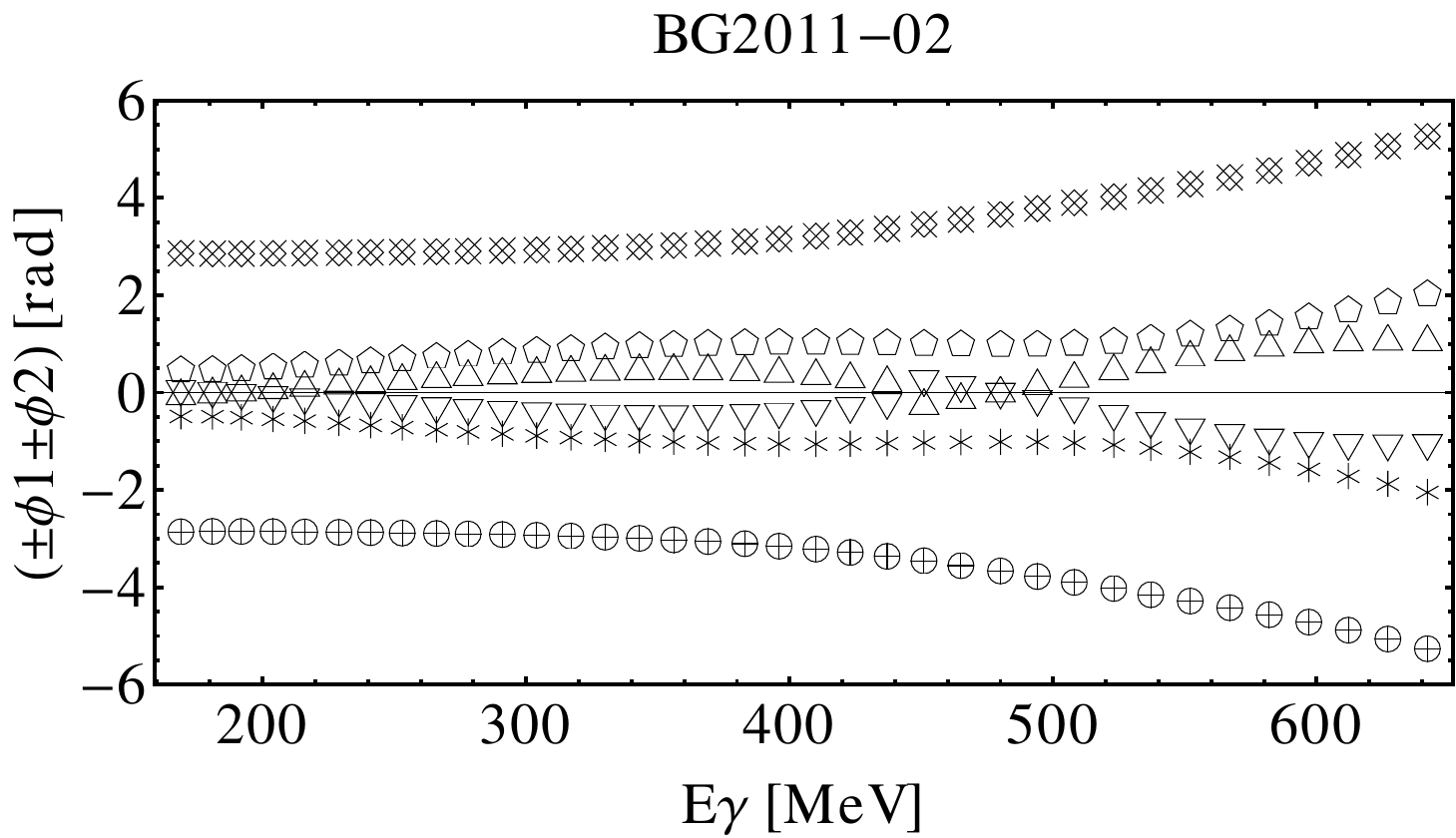}
\caption{Ambiguity diagrams for the S- and P-wave multipoles of
different partial wave analyses. The left and right panels are
obtained by using the CM12 solution of the SAID group and the
BG2011-02 solution of the Bonn-Gatchina group, respectively. The
symbols chosen are as in Fig.~\ref{fig:AmbiguityDiagramLmax1}.} \label{fig:CompareSolutions}
\end{figure}
the following minimum subsets of observables already form complete
sets that exclude the need for experimental information on recoil
polarization:
\begin{equation}
\left\{\sigma_{0}, \Sigma, T, P, F\right\}\,,\quad
\left\{\sigma_{0}, \Sigma, T, P, G\right\} \,.
\label{eq:CompleteSetsLmax1}
\end{equation}
The numerical input for the ambiguity study performed in this work
consists of a solution for multipoles given by the MAID partial wave
analysis~\cite{MAID}. As it is well known that the current
state-of-the-art partial wave analyses show quite some
deviations~\cite{Anisovich:2009zy} already for S- and P- wave
multipoles, it is interesting to compare the ambiguity diagrams for
different solutions. Fig.~\ref{fig:CompareSolutions} shows the
diagrams obtained from multipoles of the SAID group~\cite{SAID} as
well as of the Bonn-Gatchina group~\cite{BoGa}.

For all three partial wave analyses, the diagrams show a similar
structure. Symbols referring to the starting solution as well as the
double ambiguity in each case inhabit the same areas in the plot.
The most visible differences are seen in the closeness of the
symbols defining the possible accidental ambiguities at lower
energies as well as the possible appearance of intersections for
higher energies. At low energies, symbols are most nearby for the
MAID2007 solution, for which the corresponding ambiguities have
already been ruled out. Therefore it is expected that any possible
accidental ambiguities are also negligible at low energies for the
SAID and BnGa solutions. This comparison of different partial wave
analyses concludes the discussion on the S- and P-wave truncation in
this section.

\section{Conclusion and Outlook} \label{sec:ConclusionAndOutlook}
This work contains a treatment of the ambiguity problem that arises
in the truncated partial wave analysis of pseudoscalar meson
photoproduction in a consideration of single channels that have
highly suppressed t-channel exchanges. For this purpose, the
approach of Omelaenko from 1981~\cite{omel} was revisited and
supplemented by more information on intermediate calculational
steps. This above mentioned approach consists of first searching for
all possible ambiguities of the group S observables and then
selecting appropriate double polarization measurements that can
remove all additional solutions. One ambiguity, called the double
ambiguity, can be predicted just by the formalism. It can be removed
for all energy regions and all orders in the truncation angular
momentum $L$ by a measurement of the observables $G$ and $F$ or any
beam-recoil as well as target-recoil double polarization observable.
However there can also exist numerically accidental ambiguities that may require
information on additional double polarization observables.

As a numerical application of the presented formalism, the
investigation of an S- and P-wave truncation (i.e. $L=1$) also
executed similarly in Ref.~\cite{omel} was done using multipoles of
the partial wave analysis solution MAID2007~\cite{MAID} as input. It
was found that for this situation, i.e. in a treatment that
disregards measurement uncertainty, accidental ambiguities can be
neglected and only the double ambiguity has to be removed. Therefore
in this case the sets of $5$ observables
\begin{equation} \left\{\sigma_{0}, \Sigma, T, P, F\right\}
\,,\quad \left\{\sigma_{0}, \Sigma, T, P, G\right\}  \nonumber
\end{equation}
can be postulated as complete sets of observables for this simplest
case in the context of the study. As derived in
Sect.~\ref{sec:DoublePolarizationObs}, the double polarization
observables $F$ or $G$ can also be replaced by any one of the recoil
observables of the groups BR and TR.

The development of the situation for increasing $L$ is as follows.
The number of new sets of potentially ambiguous solutions is
$2^{4L}$ for every $L$. Although not all of these solutions have to
fulfill all of the consistency requirements in order to be regarded
as realistic ambiguities, the number of candidates that potentially
could fulfill all those requirements is vastly increasing. This
increasing difficulty with growing angular momentum $L$ is also
described in Ref.~\cite{omel}. It is therefore likely that, at least
as soon as real data are fitted, the complete sets given above have
to be extended by additional observables for higher values of $L$.

As an outlook it is interesting whether the results found in this
work apply to the numerical fitting of data. The following procedure
is proposed for these fits. First, numerical precision data for
polarization observables generated by use of existing PWA solutions
should be fitted. These data do not carry statistical fluctuations
and have numerical uncertainties given by the number of digits in
the tables. In this case it is expected that the accidental
ambiguities are not significant, since only precise equalities of
phases are relevant, which are relatively infrequent. The numerical
precision data could then be used in order to generate pseudo data
that are closer to the realistic situation by carrying adjustable
uncertainties~\cite{Workman:2011hi}. Fits to these data then have to
show how significant the impact of varying uncertainties is on the
appearance of additional ambiguous solutions. However, both fitting
procedures proposed until now are only preparatory steps. The final
goal is to investigate the fitting to real data from the world
database of a specific photoproduction channel, for example $\gamma
p \rightarrow \pi^{0} p$.

It remains to be seen whether it will be possible to arrive at a
final unique multipole solution by using only group S and
beam-target double polarization observables, exclusively.

\begin{acknowledgments}
   This work was supported by the Deutsche Forschungsgemeinschaft (SFB/TR16 and SFB 1044)
   and the European Community-Research Infrastructure Activity (FP7).
   The authors would like to thank all members of the Bonn-Gatchina group for fruitful
   discussions and Annika Thiel for the careful reading of the manuscript.
\end{acknowledgments}

\appendix

\section{Derivation of explicit expressions for angular polynomials} \label{sec:AppendixA}

The multipole expansion of Eqs.~(\ref{eq:MultExpF1}) to
(\ref{eq:MultExpF4}) can be written in a more convenient form for a
truncation at finite $L$
\begin{align}
F_{1} \left( W, \theta \right) &= \sum \limits_{\ell = 0}^{L} \Big\{
f_{\ell}^{(1)} \left(W\right) P_{\ell+1}^{'} \left( x \right)
 + f_{\ell}^{(2)} \left(W\right) P_{\ell-1}^{'} \left( x \right) \Big\} \mathrm{,} \label{eq:MultExpF1Rewritten} \\
F_{2} \left( W, \theta \right) &= \sum \limits_{\ell = 1}^{L} f_{\ell}^{(3)} \left(W\right) P_{\ell}^{'} \left( x \right) \mathrm{,} \label{eq:MultExpF2Rewritten} \\
F_{3} \left( W, \theta \right) &= \sum \limits_{\ell = 1}^{L} \Big\{
f_{\ell}^{(4)} \left(W\right) P_{\ell+1}^{''} \left( x \right)
 + f_{\ell}^{(5)} \left(W\right) P_{\ell-1}^{''} \left( x \right) \big\} \mathrm{,} \label{eq:MultExpF3Rewritten} \\
F_{4} \left( W, \theta \right) &= \sum \limits_{\ell = 2}^{L} f_{\ell}^{(6)} \left(W\right) P_{\ell}^{''} \left( x \right) \mathrm{,} \label{eq:MultExpF4Rewritten}
\end{align}
with $x=\cos \theta$ and the following six energy dependent
functions
\begin{align}
f_{\ell}^{(1)} \left(W\right) &= \ell M_{\ell+} \left(W\right) + E_{\ell+} \left(W\right) \mathrm{,} \hspace*{8pt} \label{eq:f1Notation11} \\
f_{\ell}^{(2)} \left(W\right) &= (\ell+1) M_{\ell-} \left(W\right) + E_{\ell-} \left(W\right) \mathrm{,} \label{eq:flNotation1} \\
f_{\ell}^{(3)} \left(W\right) &= (\ell+1) M_{\ell+} \left(W\right) + \ell M_{\ell-} \left(W\right) \mathrm{,} \label{eq:f1Notation22} \\
f_{\ell}^{(4)} \left(W\right) &= E_{\ell+} \left(W\right) -  M_{\ell+} \left(W\right) \mathrm{,} \label{eq:flNotation2} \\
f_{\ell}^{(5)} \left(W\right) &= E_{\ell-} \left(W\right) + M_{\ell-} \left(W\right) \mathrm{,} \label{eq:flNotation33} \\
f_{\ell}^{(6)} \left(W\right) &= M_{\ell+} \left(W\right) -
E_{\ell+} \left(W\right) - M_{\ell-} \left(W\right) -
E_{\ell-}\left(W\right) \,. \label{eq:flNotation3}
\end{align}
It is useful to introduce the Pochhammer symbols~\cite{Gersten}
\begin{align}
(a)_{m} := a (a+1) \ldots (a+m-1), \quad (a)_{0} := 1 \,.
\label{eq:DefPochhammers1}
\end{align}
For the special cases $(a)_{1}$ and $(1)_{m}$ this definition yields
\begin{align}
(a)_{1} = a, \quad (1)_{m} = m! \,. \label{eq:DefPochhammers2}
\end{align}
The symbols $(a)_{m}$ appear in the expansion of the hypergeometric
function~\cite{Gersten,Abramowitz}
\begin{equation}
_{2}F_{1} \left( a, b; c; Z \right) := \sum_{m = 0}^{\infty}
\frac{(a)_{m} (b)_{m}}{(c)_{m} m!} Z^{m} \,,
\label{eq:DefHypFunction}
\end{equation}
for real quantities $a$, $b$, $c$ and a generally complex argument
$Z \in \mathbb{C}$. Equation (\ref{eq:DefHypFunction}) corresponds
to a particular choice of indices in the definition of the
generalized hypergeometric function
\begin{align}
&_{n}F_{m} \left( a_{1},\ldots,a_{n}; b_{1},\ldots,b_{m}; Z \right)
:= \sum_{k = 0}^{\infty} \frac{(a_{1})_{k} \ldots
(a_{n})_{k}}{(b_{1})_{k} \ldots (b_{m})_{k} k!} Z^{k} \mathrm{.}
\label{eq:DefGeneralHypFunction}
\end{align}
It is important to note that the Legendre polynomials $P_{\ell}
\left( \cos \theta \right)$ can be expressed in terms of
hypergeometric functions, i.e.~\cite{Gersten}
\begin{equation}
P_{\ell} \left( \cos \theta \right) = \,_{2}F_{1}
\left( -\ell, \ell+1; 1; \frac{1 - c}{2} \right) \mathrm{,}
\label{eq:PlCosInTermsOfHypCos}
\end{equation}
where on the right hand side the abbreviation $c = \cos \theta$ was
chosen in the argument of $\,_{2}F_{1}$. This work features an
exchange of the angular variable $c = \cos \theta$ for $t = \tan
\theta/2$. Equation~(\ref{eq:PlCosInTermsOfHypCos}), with right hand
side rewritten in terms of $t$ takes the form~\cite{Gersten}
\begin{equation}
P_{\ell} \left( \cos \theta \right) = (1 + t^{2})^{-\ell} \,_{2}F_{1}
\left( -\ell, -\ell; 1; - t^{2} \right) \mathrm{.}
\label{eq:PlCosInTermsOfHypTan}
\end{equation}
The idea is to rewrite all derivatives of Legendre polynomials
appearing in Eqs.~(\ref{eq:MultExpF1Rewritten}) to
(\ref{eq:MultExpF4Rewritten}) in terms of hypergeometric functions
$\,_{2}F_{1}$ depending on $t$. In order to do this, a relation is
needed that can be inferred from equation (15.2.7) of
Ref.~\cite{Abramowitz}
\begin{equation}
\frac{d}{d Z} \left[ \left( 1 - Z \right)^{a} \,_{2}F_{1} \left( a,
b; c; Z \right) \right] = (-) \frac{a (c - b)}{c} \left( 1 - Z
\right)^{a - 1} \times\,_{2}F_{1} \left( a+1, b; c+1; Z \right) \,.
\label{eq:Abramowitzidentity}
\end{equation}
This identity is necessary for the determination of the derivative
of $P_{\ell} \left( \cos \theta \right)$. The first order derivative
$P_{\ell}^{\prime} \left( \cos \theta \right)$ can be rearranged as
\begin{align}
P_{\ell}^{\prime} \left( \cos \theta \right) &= \frac{d}{d \cos \theta} P_{\ell} \left( \cos \theta \right) \nonumber \\
&= \frac{d}{d \cos \theta} \left[ (1 + t^{2})^{-\ell} \,_{2}F_{1} \left( -\ell, -\ell; 1; - t^{2} \right) \right] \nonumber \\
 &= \frac{d}{d t^{2}} \left[ (1 + t^{2})^{-\ell} \,_{2}F_{1} \left( -\ell, -\ell; 1; - t^{2} \right) \right]
\times \frac{d t^{2}}{d \cos \theta} \,.
  \label{eq:PlPrimeRewritten}
\end{align}
Inspection of Eq.~(\ref{eq:TanInTermsOfCos}) facilitates the
evaluation of the second factor in the relation given above, i.e.
\begin{align}
\frac{d t^{2}}{d \cos \theta} &= \frac{d}{d \cos \theta} \tan^{2}
\frac{\theta}{2} = \frac{d}{d \cos \theta} \left[ \frac{1 - \cos
\theta}{1 + \cos \theta} \right]
 = - \frac{2}{\left(1 + \cos \theta\right)^{2}}\nonumber \\
 &= - \frac{1}{2} \left( 1 + t^{2} \right)^{2} \,.
 \label{eq:DTsquareByDCosTheta}
\end{align}
The identity (\ref{eq:Abramowitzidentity}) yields the first factor
on the right hand side of Eq.~(\ref{eq:PlPrimeRewritten}), so that
the final result becomes
\begin{widetext}
\begin{equation}
P_{\ell}^{\prime} \left( \cos \theta \right) = \frac{1}{2} \ell
(\ell+1) (1 + t^{2})^{-\ell+1} \,_{2}F_{1} \left( -\ell+1, -\ell; 2;
- t^{2} \right) \,.
\label{eq:DerivativePlCosInTermsOfHypTan}
\end{equation}
\end{widetext}
The same procedure also yields an expression for the second
derivative of $P_{\ell} \left( \cos \theta \right)$
\begin{widetext}
\begin{equation}
P_{\ell}^{\prime \prime} \left( \cos \theta \right) = \frac{1}{8}
(\ell-1) \ell (\ell+1) (\ell+2) (1 + t^{2})^{-\ell+2} \,_{2}F_{1}
\left( -\ell+2, -\ell; 3; - t^{2} \right) \,.
\label{eq:2ndDerivativePlCosInTermsOfHypTan}
\end{equation}
\end{widetext}
%
Everything assembled until now facilitates the evaluation of the
polynomial $A_{2L}^{\prime} \left(t\right)$ that appears in the
amplitude $b_{4}$ of Eq.~(\ref{eq:b4Aprime}). First of all, the term
%
$\left[ F_{1} \left(\theta\right) - \left( \cos \theta - i \sin \theta \right) F_{2} \left(\theta\right) \right] \label{eq:b4CalcTerm}$
%
that can be deduced from Eq.~(\ref{eq:b4BasicForm}), when written in terms of the variable
$t$ reads (see Eq.~(\ref{eq:SinCosInTermsOfTan}))
\begin{equation}
\left[ F_{1} \left(\theta\right) + \frac{1}{(1 + t^{2})} (t + i)^{2} F_{2} \left(\theta\right) \right] \mathrm{.} \label{eq:b4CalcTermInTermsOfTan}
\end{equation}
Insertion of the multipole expansions (\ref{eq:MultExpF1Rewritten})
and (\ref{eq:MultExpF2Rewritten}) yields
\begin{align}
\sum_{\ell = 0}^{L} \Bigg[ &f_{\ell}^{(1)} P_{\ell+1}^{\prime}
\left( \cos \theta \right) + f_{\ell}^{(2)} P_{\ell-1}^{\prime}
\left( \cos \theta \right) + \frac{(t + i)^{2}}{(1 + t^{2})}
f_{\ell}^{(3)} P_{\ell}^{\prime} \left( \cos \theta \right) \Bigg]
\,.
\label{eq:b4CalcTermTanMultExpInserted}
\end{align}
Usage of (\ref{eq:DerivativePlCosInTermsOfHypTan}) and pulling out
an overall factor $(1 + t^{2})^{-L}$ out of the sum already gives
the result for $b_{4}$ given in the main text
\begin{widetext}
\begin{align}
b_{4} \left(\theta\right) &= \frac{\mathcal{C}}{4} \frac{\exp \left[ i \theta/2 \right]}{(1 + t^{2})^{L}} \sum_{\ell = 0}^{L} \Big\{ f_{\ell}^{(1)} (\ell+1) (\ell+2) (1 + t^{2})^{L-\ell} \,_{2}F_{1} \left( -\ell, -\ell-1; 2; - t^{2} \right) \nonumber \\
 &  + f_{\ell}^{(2)} \ell (\ell-1) (1 + t^{2})^{L-\ell+2} \,_{2}F_{1} \left( -\ell+2, -\ell+1; 2; - t^{2} \right) \nonumber \\
 &  + f_{\ell}^{(3)} \ell (\ell+1) (t + i)^{2} (1 + t^{2})^{L-\ell} \,_{2}F_{1} \left( -\ell+1, -\ell; 2; - t^{2} \right) \Big\} \mathrm{.} \label{eq:b4WithHypFunctions}
\end{align}
\end{widetext}
In order to determine the polynomial $B_{2L}^{\prime} \left(t\right)
= A_{2L}^{\prime} \left(t\right) + t D_{2L - 2}^{\prime}
\left(t\right)$ of the amplitude $b_{2}$ of
Eq.~(\ref{eq:b2AprimeDprime}), it is sufficient to infer the form of
$D_{2L-2}^{\prime} \left(t\right)$ by inspection of the formula
(\ref{eq:b2BasicForm}). It is therefore necessary to rewrite the term
\begin{equation}
i \sin \theta \left[ F_{3} \left(\theta\right) + \left( \cos \theta - i \sin \theta \right) F_{4} \left(\theta\right) \right] \mathrm{,} \label{eq:b2CalcTerm}
\end{equation}
in terms of the variable $t$
\begin{equation}
\frac{2 i t}{(1 + t^{2})} \left[ F_{3} \left(\theta\right) - \frac{1}{(1 + t^{2})} (t + i)^{2} F_{4} \left(\theta\right) \right] \mathrm{.} \label{eq:b2CalcTermInTermsOfTan}
\end{equation}
Invoking the multipole expansions (\ref{eq:MultExpF3Rewritten}) and (\ref{eq:MultExpF4Rewritten}) yields
\begin{align}
\frac{2 i t}{(1 + t^{2})} \sum_{\ell=0}^{L} \Bigg[ &f_{\ell}^{(4)}
P_{\ell+1}^{\prime \prime} \left( \cos \theta \right) +
f_{\ell}^{(5)} P_{\ell-1}^{\prime \prime} \left( \cos \theta \right)
- \frac{(t + i)^{2}}{(1 + t^{2})} f_{\ell}^{(6)} P_{\ell}^{\prime
\prime} \left( \cos \theta \right) \Bigg] \,.
\label{eq:eq:b2CalcTermTanMultExpInserted}
\end{align}
Usage of (\ref{eq:2ndDerivativePlCosInTermsOfHypTan}) in a similar
way yields the expression for $D_{2L-2}^{\prime} \left(t\right)$
that is already given in Eq.~(\ref{eq:b2AprimeDprime}) of the main
text,
%
\begin{widetext}
\begin{align}
D_{2L-2}^{\prime} \left(t\right) &= \frac{1}{4} \sum_{\ell = 0}^{L} \Big\{ (if_{\ell}^{(4)}) \ell (\ell+1) (\ell+2) (\ell+3) (1 + t^{2})^{L-\ell} \,_{2}F_{1} \left( -\ell+1, -\ell-1; 3; - t^{2} \right) \nonumber \\
 & \hspace*{4pt} + (if_{\ell}^{(5)}) (\ell-2) (\ell-1) \ell (\ell+1) (1 + t^{2})^{L-\ell+2} \,_{2}F_{1} \left( -\ell+3, -\ell+1; 3; - t^{2} \right) \nonumber \\
 & \hspace*{4pt}  - (if_{\ell}^{(6)}) (\ell-1) \ell (\ell+1) (\ell+2) (t + i)^{2} (1 + t^{2})^{L-\ell} \,_{2}F_{1} \left( -\ell+2, -\ell; 3; - t^{2} \right) \Big\} \mathrm{.} \label{eq:D2PrimeHypFunctions}
\end{align}
\end{widetext}
Furthermore, the expressions for $A_{2L}^{\prime} \left(t\right)$
and $B_{2L}^{\prime} \left(t\right)$ given in this appendix can be
further simplified and be brought into the form
\begin{align}
A_{2L}^{\prime} \left(t\right) &= \sum_{\ell = 0}^{2L} a_{\ell} t^{\ell} \mathrm{,} \label{eq:APrimePolynomialformAppendix} \\
B_{2L}^{\prime} \left(t\right) &= \sum_{\ell = 0}^{2L} b_{\ell} t^{\ell} \mathrm{,} \label{eq:BPrimePolynomialformAppendix}
\end{align}
with explicit formulae for the complex expansion coefficients
$a_{\ell}$ and $b_{\ell}$ in terms of multipoles (see
Ref.~\cite{Gersten}, where similar expressions are given for $\pi N$
scattering).

\section{Linear relations among $\left\{a_{i}, \hspace*{2pt} b_{i}\right\}$
and $\left\{E_{\ell\pm}, \hspace*{2pt} M_{\ell\pm}\right\}$ for $L=1$ and $L=2$}
\label{sec:AppendixC}

Linear relations among multipoles and complex polynomial coefficients for $L=1$:
\begin{equation}
\left[ \begin{array}{c}
        E_{0+} \\
        E_{1+} \\
        M_{1+} \\
        M_{1-}
       \end{array} \right] = \frac{a_{2}}{2} \left[ \begin{array}{cccc}
                                                    1 & 1 & 0 & 0 \\
                                                    -\frac{1}{6} & \frac{1}{6} & 0 & -\frac{i}{6} \\
                                                    -\frac{1}{6} & \frac{1}{6} & -\frac{i}{3} & \frac{i}{6} \\
                                                    \frac{1}{3} & -\frac{1}{3} & -\frac{i}{3} & -\frac{i}{3} \end{array} \right] \left[ \begin{array}{c}
        1 \\
        \hat{a}_{0} \\
        \hat{a}_{1} \\
        \hat{b}_{1}
       \end{array} \right] \mathrm{.} \label{eq:MatrixEquationLmax1}
\end{equation}
Similar relations for the case $L=2$:
%
\begin{widetext}
\begin{equation}
\left[ \begin{array}{c}
        E_{0+} \\
        E_{1+} \\
        M_{1+} \\
        M_{1-} \\
        E_{2+} \\
        E_{2-} \\
        M_{2+} \\
        M_{2-} \\
       \end{array} \right] = \frac{a_{4}}{2} \left[ \begin{array}{cccccccc}
                                                    \frac{2}{3} & \frac{2}{3} & 0 & \frac{1}{6} & 0 & 0 & \frac{1}{6} & 0 \\
                                                    -\frac{1}{6} & \frac{1}{6} & 0 & 0 & 0 & -\frac{i}{12} & 0 & -\frac{i}{12} \\
                                                    -\frac{1}{6} & \frac{1}{6} & -\frac{i}{6} & 0 & -\frac{i}{6} & \frac{i}{12} & 0 & \frac{i}{12} \\
                                                    \frac{1}{3} & -\frac{1}{3} & -\frac{i}{6} & 0 & -\frac{i}{6} & -\frac{i}{6} & 0 & -\frac{i}{6} \\
                                                     \frac{1}{45} & \frac{1}{45} & 0 & 0 & 0 & -\frac{i}{45} & -\frac{1}{45} & \frac{i}{45} \\
                                                     \frac{1}{30} & \frac{1}{30} & 0 & \frac{1}{12} & 0 & \frac{i}{20} & -\frac{7}{60} & -\frac{i}{20} \\
                                                     \frac{1}{45} & \frac{1}{45} & -\frac{i}{30} & -\frac{1}{30} & \frac{i}{30} & \frac{i}{90} & \frac{1}{90} & -\frac{i}{90} \\
                                                      -\frac{1}{30} & -\frac{1}{30} & -\frac{i}{30} & \frac{1}{20} & \frac{i}{30} & -\frac{i}{60} & -\frac{1}{60} & \frac{i}{60} \\  \end{array} \right] \left[ \begin{array}{c}
        1 \\
        \hat{a}_{0} \\
        \hat{a}_{1} \\
        \hat{a}_{2} \\
        \hat{a}_{3} \\
        \hat{b}_{1} \\
        \hat{b}_{2} \\
        \hat{b}_{3}
       \end{array} \right] \mathrm{.} \label{eq:MatrixEquationLmax2}
\end{equation}
\end{widetext}


\begin{thebibliography}{99}

\bibitem{Beringer:1900zz}J.~Beringer {\it et al.}  [Particle Data Group Collaboration],
  Phys.\ Rev.\ D {\bf 86}, 010001 (2012).

\bibitem{Anisovich:2011fc}
  A.~V.~Anisovich, R.~Beck, E.~Klempt, V.~A.~Nikonov, A.~V.~Sarantsev and U.~Thoma,
  Eur.\ Phys.\ J.\ A {\bf 48}, 15 (2012).

\bibitem{Hoehler84} G. Hoehler, \emph{Pion Nucleon Scattering}, Part 2, Landolt-Bornstein:
Elastic and Charge Exchange Scattering of Elementary Particles, Vol.
9b (Springer-Verlag, Berlin, 1983).

\bibitem{CMB} R. E. Cutkosky, C. P. Forsyth, R. E. Hendrick, and R. L. Kelly, Phys. Rev. \textbf{D 20}, 2839 (1979).

\bibitem{Arndt:2006bf}
  R.~A.~Arndt, W.~J.~Briscoe, I.~I.~Strakovsky and R.~L.~Workman,
  Phys.\ Rev.\ C {\bf 74}, 045205 (2006).

\bibitem{Drechsel:2007if}
  D.~Drechsel, S.~S.~Kamalov and L.~Tiator,
  Eur.\ Phys.\ J.\ A {\bf 34}, 69 (2007).

\bibitem{Chen:2007cy}
  G.~Y.~Chen, S.~S.~Kamalov, S.~N.~Yang, D.~Drechsel and L.~Tiator,
  Phys.\ Rev.\ C {\bf 76}, 035206 (2007).

\bibitem{Ronchen:2012eg}
  D.~Roenchen, M.~Doring, F.~Huang, H.~Haberzettl, J.~Haidenbauer, C.~Hanhart, S.~Krewald and U.~-G.~Meissner {\it et al.},
  Eur.\ Phys.\ J.\ A {\bf 49}, 44 (2013).

\bibitem{Kamano:2013iva}
  H.~Kamano, S.~X.~Nakamura, T.~-S.~H.~Lee and T.~Sato,
  Phys.\ Rev.\ C {\bf 88}, 035209 (2013).

\bibitem{Shrestha:2012ep}
  M.~Shrestha and D.~M.~Manley,
  Phys.\ Rev.\ C {\bf 86}, 055203 (2012).

\bibitem{Shklyar:2012js}
  V.~Shklyar, H.~Lenske and U.~Mosel,
  Phys.\ Rev.\ C {\bf 87}, 015201 (2013).

\bibitem{Barker75} I.~S. Barker, A. Donnachie, J.~K. Storrow, Nucl. Phys. B {\bf 95}, 347 (1975).

\bibitem{Keaton:1996pe}G.~Keaton and R.~Workman, Phys.\ Rev.\  C {\bf 54}, 1437 (1996).

\bibitem{chiang}W.-T. Chiang and F. Tabakin, Phys. Rev. C {\bf 55}, 2054 (1997).

\bibitem{Ireland:2010bi}D.~G.~Ireland, Phys.\ Rev.\ C {\bf 82}, 025204
(2010).

\bibitem{Workman:2011hi}R.~L.~Workman, M.~W.~Paris, W.~J.~Briscoe, L.~Tiator, S.~Schumann, M.~Ostrick,
   S.~S.~Kamalov, Eur.\ Phys.\ J.\ A {\bf 47}, 143 (2011).

\bibitem{Sandorfi:2010uv}A.~M.~Sandorfi, S.~Hoblit, H.~Kamano, T.~-S.~H.~Lee,
  J.\ Phys.\ G {\bf 38}, 053001 (2011).

\bibitem{Vrancx:2013pza}T.~Vrancx, J.~Ryckebusch, T.~Van Cuyck and P.~Vancraeyveld,
  Phys.\ Rev.\ C {\bf 87}, 055205 (2013).

\bibitem{Sikora:2013vfa}
  M.~H.~Sikora, D.~P.~Watts, D.~I.~Glazier, P.~Aguar-Bartolome, L.~K.~Akasoy, J.~R.~M.~Annand, H.~J.~Arends and K.~Bantawa {\it et al.},
  [arXiv:1309.7897 [nucl-ex]].

\bibitem{Goldberger:1963} M.~L.~Goldberger, H.~W.~Lewis and K.~M.~Watson,
  Phys.\ Rev.\  {\bf 132}, 2764 (1963).

\bibitem{Ivanov:2012na}I.~P.~Ivanov, Phys.\ Rev.\ D {\bf 85}, 076001 (2012).

\bibitem{Tiator:2011tu}
  L.~Tiator, AIP Conf.\ Proc.\  {\bf 1432}, 162 (2012)  [arXiv:1109.0608 [nucl-th]].

\bibitem{Tiator:2011a}L.~Tiator, (Bled Workshops in Physics. Vol. 13 No. 1)  [arXiv:1211.3927 [nucl-th]].

\bibitem{omel}A.~S. Omelaenko, Sov. J. Nucl. Phys. {\bf 34}, 406 (1981).

\bibitem{grushin} V.~F. Grushin, in {\it Photoproduction of Pions on Nucleons and Nuclei}, edited
by A.~A. Komar (Nova Science, New York, 1989), p. 1ff.

\bibitem{MAID}
(MAID Partial Wave Analysis) http://www.kph.uni-mainz.de/MAID/

\bibitem{SAID}
(SAID Partial Wave Analysis) http://gwdac.phys.gwu.edu/

\bibitem{BoGa}
(Bonn Gatchina Partial Wave Analysis) http://pwa.hiskp.uni-bonn.de/

\bibitem{Gersten} A. Gersten, Nucl. Phys. B {\bf 12}, 537 (1969).

\bibitem{Barrelet} E. Barrelet, Nuovo Cimento {\bf 8A}, 331 (1972).

\bibitem{VanHorn} A. J. Van Horn, Nucl. Phys. B {\bf 87}, 157 (1975).

\bibitem{CGLN}
G. F. Chew, M. L. Goldberger, F. E. Low, and Y. Nambu, Phys. Rev. {\bf 106}, 1345 (1957).

\bibitem{Abramowitz}
M. Abramowitz and I.A. Stegun, Handbook of Mathematical Functions, Dover Publishing (1972).

\bibitem{FTS92} C.~G. Fasano, F. Tabakin, B. Saghai, Phys. Rev. C {\bf 46}, 2430 (1992).

\bibitem{Conventions} A. M. Sandorfi, B. Dey, A. Sarantsev, L. Tiator and R. Workman,
AIP Conf.\ Proc.\  {\bf 1432}, 219 (2012)
[arXiv:1108.5411v2[nucl-th]].

\bibitem{Anisovich:2009zy}A.~V.~Anisovich, E.~Klempt, V.~A.~Nikonov {\it et al.},
  Eur.\ Phys.\ J.\ A {\bf 44}, 203-220 (2010).


\end{thebibliography}
\end{document}